\documentclass[12pt,longbibliography,eprint]{revtex4-1}
\usepackage{epigraph}
\usepackage{graphicx}
\usepackage{epstopdf}
\usepackage{hyperref}

\begin{document}
\preprint{V.M.}
\title{Mathematical Foundations of Realtime Equity Trading.
  Liquidity Deficit and Market Dynamics.
  Automated Trading Machines.}
\author{Vladislav Gennadievich \surname{Malyshkin}} 
\email{malyshki@ton.ioffe.ru}
\affiliation{Ioffe Institute, Politekhnicheskaya 26, St Petersburg, 194021, Russia}

\author{Ray Bakhramov} 
\email{rbakhramov@forumhedge.com}
\affiliation{Forum Asset Management LLC, 733 Third Avenue,
New York, NY 10017}

\date{September 13, 2015}

\begin{abstract}
\begin{verbatim}
$Id: LD.tex,v 1.211 2016/12/04 09:32:16 mal Exp $
\end{verbatim}
We postulates, and then show experimentally,
that liquidity deficit is the driving
force of the markets. In the first part of the paper
a kinematic of liquidity deficit is developed.
The calculus-like approach, which is based on Radon--Nikodym
derivatives and their generalization,
allows us to calculate important characteristics of
observable market dynamics.
In the second part of the paper this calculus is used in
an attempt to build a dynamic equation in the form:
future price tend to the value
maximizing the number of shares traded per unit time.
To build a practical automated trading machine P\&L dynamics instead of
price dynamics is considered. This allows
a trading automate resilient to
catastrophic P\&L drains to be built.
The results are very promising,
yet when all the fees and trading commissions are taken into account,
are close to breakeven.
In the end of the paper important criteria for automated trading systems
are presented. We list the system types
that can and cannot make money on the market. These
criteria can be successfully applied not only by automated trading machines,
but also by a human trader.
\end{abstract}

\maketitle

\epigraph{Thou art wearied in the multitude of thy counsels. Let now the astrologers, the stargazers, the monthly prognosticators, stand up, and save thee from these things that shall come upon thee.}{Isa.47:13}

\section{\label{intro}Introduction}
Market dynamic study
attract a lot of attention\cite{mandelbrot2014misbehavior,corcoran2007long,mccauley2009dynamics,choi2015information,2015arXiv151005510G}.
We start with a short review about  available
 data for equity trading  market.
Exchange trading is typically consist of sending limit orders
at specific price. Depending on liquidity available
this order can be either executed (matched to an order
of opposite type), or, in case no matching liquidity
available, to be put into the order book.
This is so called double auction process (both ``buy'' and ``sell''
orders are put into the order book;
we will use NASDAQ ITCH terminology, where ``bid orders'' are called ``buy orders''
and ``offer orders'' are called ``sell orders''), the difference
between best sell and best buy orders in the order book is spread.
Our experiments show that since about 2008 order book (tested on NASDAQ
ITCH total view feed and on CME data)
carry no valuable information. Our study show that:
1) More than 90\% of orders
being at best price level at some time end being cancelled,
not executed (order-stuffing like behavior). The Ref. \onlinecite{nasdaqord}
authors came to the same conclusion regarding cancellation.
2) Spread is also very misleading indicator.
Our experiments show that a limit order being put inside spread
interval has very high chances of being immediately executed.
There are two reason for that: many market participants do not show
their liquidity if the price they can accept is inside spread interval
and ``hidden'' type of orders (the ones not being broadcast
as available in order book, but actually existing
in exchange order book.
Such ``hidden'' type orders cost more on NASDAQ. Executed hidden order id was
actually available before October 6, 2010, but after this date NASDAQ
broadcast 0 as hidden order id, see Appendix A of Ref. \onlinecite{itchfeed}).
3) There is a long discussion that order book observable spread
is actually higher that ``actual'' spread\cite{secchair}
because of order book manipulation (typically
either stuffing the book or attempting to frontrun).

Based on all the information above we state that even
for a hedge fund order book information is incomplete or
manipulative\cite{NoteBookInfo}.

We can imagine that to an exchange or to major brokerages
some additional information can be potentially available,
but not to general public. So we go
for a much more ambitious goal - try to predict market dynamics
based on price and volume of executed orders only.
The information of executed orders is
legally required to be available (Dark Pools and  brokerages
internal order matching can be a problem to some degree, but not much),
and it is much more costly to manipulate through actual trades.
  Whether market manipulation is possible via trade execution
  is one of most fundamental problem of market analysis.
  Naive pump-and-dump type of  manipulation (buy shares to drive market up,
  then sell then) actually never work because of
  concave type of impact\cite{farmerimpact}.
  The volume required to ``pump'' shares
  from current price to some higher price
  is greater that the volume on a way back.
  This way a ``manipulator'' would not be able
  to sell all shares bought, and to sell the remaining shares
  price should go below its initial value, thus this strategy would lose
  money.
  We disagree on this type of ``active trading strategy''
  with\cite{EyakushevComm} who observed convex demand
  on a number of low liquidity stocks.
  Our experiments on low and medium liquidity stocks
  show that in a situation when overall market is flat
  once price of some stock is driven by excessive buying to
  some level the market maker (or some other market participant)
  start buying whatever volumes is available, so
  after some price level almost no further price movement possible,
  even on very high volume. 

On NASDAQ placing a limit order and then cancelling it
cost almost nothing, what create a free opportunity
to manipulate order book. From our opinion, the most effective
way to suppress order book manipulations can be an introduction,
not an artificial delay, what HFT opponents
often propose, but order fee structure,
similar in philosophy to currently existing
execution fee structure (exchange rebate and liquidity removal fee,
but for orders exchange rebate will be zero
and liquidity removal fee will be small).
Proposed fee structure to suppress order book manipulation may be this:
\begin{itemize}
\item Your order in order book was matched by somebody else
  order -- your get ``exchange rebate'',
  a fraction of a cent per each share,
  same as it is now on most exchanges.
\item You matched somebody else order in the order book (remove liquidity)
  you are charged ``liquidity removal fee'', which is slightly greater
  than the ``exchange rebate'', same as it is now on most exchanges.
\item New fee proposed: You cancelled your own order
  in order book: you are charged ``order removal fee'',
  which should be much lower than the difference between ``liquidity removal fee'' and ``exchange rebate''
  for executed orders.
\end{itemize}
This fee structure would make order book
manipulation non--free, but in the same time it would not not suppress
actual trading (execution orders matching).

The major risk for manipulator through order execution
is not so much the fees, but market movement.
With a spread about few cents market manipulator through execution takes a huge risk of market moving
against him.
Currently only trade execution is expensive to manipulate, so
our theory uses only trade execution information:
for a company ``XYZ'' at time $t$ an order of size $v$ was executed at price $p$.
There are few other worth to mention attributes, not used
in this paper, but possessing some interesting
properties (we are going to discuss them in a separate publication).
\begin{itemize}
  \item
    Volume multiplied by spread.
  \item
    Execution type: ``sell'' (when buy order matched sell limit order in order book)
    or of type ``buy'' (when sell order matches buy limit order in order book).
  \item
    A ``signed volume'' is used by some traders\cite{EyakushevComm}: when type is ``sell'' use order size $v$,
    when type is ``buy'', use $-v$.
  \item
    Order book information, from out opinion, is only valuable\cite{NoteBookInfo}.
    as a product of (possibly signed) order size multiplied by $\tau_{oe}$,
    the difference  between execution time and limit order origination time.
    An important property of this attribute $\tau_{oe} dv/dt$
    is that it combines
    the characteristics of original limit order $\tau_{oe}$
    and matching to it marker order $dv/dt$ (execution flow),
    thus the attribute can be considered(when signed volume is used)
    as proportional to supply-demand disbalance.
\end{itemize}

\section{\label{kin}Kinematics}

Executed orders is a set of timeseries observations.
We convert observations data from timeseries space to an invariant basis
space. Selection of the basis depends on a number of factors.
The simplest selection is polynomials basis $Q_k(x)$, where $Q_k$ is a polynomial
of a degree $k$,
with some measure
selected to define inner product.
The three bases below are the most convenient ones to transform
a timeserie $f(t_i)$ to moment $f_k$ space.

\noindent
Laguerre basis:
\begin{eqnarray}
x&=&t/\tau \\
f_k&=&\int\limits_{-\infty}^{0} Q_k(x) f(t) \exp(x) dx
\label{flaguerre} \\
d\mu&=&\exp(x) dx \label{muflaguerre} \\
\mathrm{supp}(\mu(x)) &=& x \in [-\infty , 0] \label{lagurrerange}
\end{eqnarray}

\noindent
Shifted Legendre basis:
\begin{eqnarray}
  x&=&\exp(t/\tau) \\
  f_k&=&\int\limits_{-\infty}^{0} Q_k(x) f(t) \exp(t/\tau) dt/\tau 
  = \int\limits_{0}^{1} Q_k(x) f(t) dx \label{slegendre} \\
  d\mu&=& \exp(t/\tau) dt/\tau   = dx \label{muslegendre}\\
  \mathrm{supp}(\mu(x)) &=& x \in [0 , 1] \label{shiftedlegendrerange}
\end{eqnarray}

\noindent
Price basis:
\begin{eqnarray}
  x&=&p \\
  f_k&=&\int\limits_{-\infty}^{0} Q_k(p(t)) f(t) \exp(t/\tau) dt/\tau  \label{pspace}\\
  d\mu&=&\exp(t/\tau) dt/\tau  \label{mupspace} \\
\mathrm{supp}(\mu(p(t))) &=& t \in [-\infty , 0] 
\end{eqnarray}

\noindent
Any timeserie $f(t_i)$ with thousands (and even millions) of observations
can be converted to a limited number of basis moments $f_k$.
The $0$-th moment $f_0$ is exponential moving average of $f$ with timescale $\tau$.
For our theory we need large number of moments, typically at least a dozen,
what create numerical instability if $f_k$ are naively calculated.
For three bases above the basis functions are polynomials but the measure is different: (\ref{muflaguerre}) , (\ref{muslegendre}), (\ref{mupspace}).
Let us define average symbols $<>$ (bra-ket quantum mechanics notations)
as an integral over measure support:
\begin{eqnarray}
  <g>_{\mu}&=&\int\limits_{\mathrm{supp}(\mu)} g d\mu
  \label{innerprod}
\end{eqnarray}
All the results are invariant
with respect to polynomial $Q_k$ choice as long it is
of $k$-th order, e.g. monomials can be used $Q_k=x^k$.
But this naive style of basis selection causes severe numerical instability
at large $k$, typically for  all $k>5$.
The specific basis selection is a very delicate
question\cite{gautschi2004orthogonal}
 which we discuss
 briefly in Appendix \ref{basisses} 
and the problem to be discussed in details in a separate publication.
Short result : for numerical stability the basis $Q_k(x)$ should
be selected in a way $Q_k(x)$ are orthogonal with respect
to some positive measure, e.g. $d\mu(x)$.
The simplest $Q_k(x)$ choice 
is orthogonal polynomials
with respect to measure $d\mu(x)$.

For Laguerre basis (\ref{muflaguerre})
selection  $Q_k(x)=L_k(-x)$, where $L_k(x)$
are Laguerre polynomials, make basis orthogonal
$\left<Q_i Q_j\right>_{\mu}=\delta_{ij}$.
For Shifted Legendre basis (\ref{muslegendre})
selection $Q_k(x)=P_k(2x-1)$, where $P_k(x)$
are Legendre polynomials  make basis orthogonal
$\left<Q_i Q_j\right>_{\mu}=\frac{1}{2i+1}\delta_{ij}$.
For Price basis (\ref{mupspace})
the orthogonal polynomials are non--classic,
but selection of monomials $Q_k(x)=(p-p^*)^k$
or Hermite polynomials $Q_k(x)=H_k(\frac{p-p^*}{\sigma})$
often give good enough numerical stability for not very large $k$
(Here $p^*$ is some price close to average and
$\sigma$ is some value close to standard deviation;
again, the result does not depend on selection of $p^*$ or $\sigma$,
only numerical stability of calculations may depend on these values.)
The (\ref{mupspace}) basis is very convenient for quasi--stationary
consideration of market dynamics
\cite{2016arXiv160204423G}.
However, for time--dependent dynamics it requires the $dp/dt$ moments,
that carry much less information than the $v$, $dv/dt$ and $d^2v/dt^2$ moments.
In the bases (\ref{muflaguerre}) and (\ref{muslegendre})
the $v$  and $d^2v/dt^2$ moments can be easily calculated from the $dv/dt$ moments
using integration by parts.

Before we go further,
let us show some familiar calculations using the basis we introduced.

1.Interpolate price (assume $f=p$) by a polynomial of $n$-th order
using least squares approximation.
\begin{eqnarray}
  &&\left<\left[f-\sum\limits_{s=0}^{s=n} \alpha_s Q_s(x)\right]^2\right>_{\mu}\to \min \\
  &&G_{kl}=<Q_k(x)Q_l(x)>_{\mu}\\
  &&f(x)=\sum_{k,l=0}^{k,l=n}Q_k(x)G^{-1}_{kl}\left<Q_l(x)f\right>_{\mu}
  \label{pinterp}
\end{eqnarray}
Here $G^{-1}$ is a matrix inverse to Gramm matrix $G$
calculated in basis $Q_k$ using inner product (\ref{innerprod}).
The interpolation polynomial (\ref{pinterp}) is of $n$-th order and depend
on the moments $\left<Q_k(x)f\right>$, where $k=0..n$
(In Ref. \onlinecite{gautschi2004orthogonal}  only
monomials moments $\left<x^kf\right>$ are called ``moments''
and the $\left<Q_k(x)f\right>$
are called ``modified moments'', we would call all of them ``moments'').
Note that interpolation polynomial typically give
good interpolation in the middle of interval, but exhibit
oscillations near interval ends (Runge oscillations).

2. Given two prices $p$ and $q$ calculate covariance between then.
\begin{eqnarray}
  &&\overline{(p-\overline{p})(q-\overline{q})}= \\
  &=&\sum_{k,l=0}^{k,l=n}\left<Q_l(x)q\right>_{\mu}G^{-1}_{kl}\left<Q_l(x)p\right>_{\mu}-<Q_0p>_{\mu}<Q_0q>_{\mu} \label{covar} \\
  &&\overline{p}=<Q_0 p>_{\mu} \\
  &&\overline{q}=<Q_0 q>_{\mu}
\end{eqnarray}
The measure $\mu$ (\ref{innerprod}) in general case is not necessary normalized
to 1 and because of this all
averages in (\ref{covar}) should be divided by
the normalizing factor $<Q_0>$ equal to an integral from a constant $Q_0$,
but all three measures we consider have $<1>=1$, so if $Q_0=1$
this normalizing factor can be omitted, see the see Appendix \ref{spur} for exact formulas in general case.
Note that (\ref{covar}) provide very efficient(linear time) algorithm 
of stock prices cross-correlation calculation.
For every stock calculate $[0..n]$ moments $\left<Q_k(x)p\right>$
forming a vector,
then obtain covariance through scalar product of these
vectors with Gramm matrix inverse used as a scalar product matrix
(note here, that if original $Q_k$ basis is orthogonal then
Gramm matrix is diagonal and its inversion process is stable,
while in general case the process of Gramm matrix inversion is numerically unstable\cite{malha}).

\subsection{\label{radinnyk}Radon--Nikodym derivatives and rational approximation}
Consider reproducing kernel $K(x,y,\mu)$ for a positive measure $d\mu$ 
\begin{eqnarray}
  M_{\mu; ij}[f]&=&<Q_i f Q_j>_{\mu} \label{mc} \\
  K(x,y,\mu)&=&Q_i(x)\left(M_{\mu}[1]\right)_{ij}^{-1} Q_j(y)
  \label{Krepr} \\
  K(x,y,\mu)&=&\mathbf{Q}(x)G_{\mu}^{-1}\mathbf{Q}(y) \label{vKrepr}
\end{eqnarray}
(Here and below we assume
a summation $[0..n]$ over ``silent'' indexes $i,j$. Another notation
we will use from time to time is vector notation, where bold $\mathbf{Q}$
define the entire vector $Q_k$ and matrix indexes are omitted.
The Eq. (\ref{vKrepr}) is exactly the same as (\ref{Krepr})
but written in vector notation.)
For arbitrary $P(y)=\alpha_i Q_i(y)$ (\ref{Krepr}) gives
$P(x)=\left<K(x,y,\mu)P(y)\right>_{\mu(y)}$. The $1/K(x,x,\mu)$
is a Christoffel function, related to the ``density'' of measure $\mu$
at point $x$, for example Gaussian quadrature weights built for the measure $\mu$
are equal to exactly $1/K(x,x,\mu)$ at $x$ equal to quadrature nodes\cite{totik}.

Consider two positive measures $d\mu(x)$ and $d\nu(x)$. Their ratio $\frac{d\nu}{d\mu}$
is called Radon--Nikodym derivative\cite{kolmogorovFA} and is of extreme importance
in market analysis\cite{taleb2014silent}. The most important for us would be to
estimate shares trading rate, or executed orders flow, $I$
\begin{eqnarray}
  I&=&\frac{dv}{dt} \label{I}
\end{eqnarray}
$I$  is the number of shares traded in unit time and is always positive.
The higher $I$ is the more active trading is.

The problem is to estimate Radon--Nikodym derivative $\frac{d\nu}{d\mu}$
at $x$ given the only moments of measures $\mu$ and $\nu$. This can be estimated,
for example, through Christoffel functions ratio\cite{BarrySimon}
\begin{eqnarray}
\frac{d\nu}{d\mu}(x)&=&\frac{K(x,x,\mu)}{K(x,x,\nu)} \label{bsim}
\end{eqnarray}
The  estimation (\ref{bsim}) is a ratio of two polynomials
of order $2n$. In contrasts with least squares approximation (\ref{pinterp})
(use $f=d\nu/d\mu$ in (\ref{pinterp})),
the (\ref{bsim}) preserves sign of interpolated function
$d\nu/d\mu$, does not diverge when $x\to\infty$, it tends to a constant instead,
and does not exhibit diverging oscillations near measure support edges.
The estimation requires $0..2n$ moments to be known (instead of just $0..n$ moments for least squares approximation).
As we stated in the beginning of the Chapter \ref{kin}
numerical estimation
for large $n$ is unfeasible (because of numerical instabilities)
unless a basis $Q_s$ orthogonal with respect to some measure (not necessary the $\mu$, but $\mu$ is often good enough)
is chosen. This approach allows us to calculate Radon--Nikodym derivative
to very high $n$ (up to 15-20 in Shifted Legendre basis and up to 12-15 in Lagurre basis, Chebyshev basis sometimes allows to use $n$ up to 30).

The approximation (\ref{bsim}) requires both measures to be
positively defined. There is exist a
different numerical estimation of Radon--Nikodym derivative, requiring only one
 measure $d\mu$ to be positive:
\begin{eqnarray}
\frac{d\nu}{d\mu}(x)&=&\frac{Q_i(x)\left(M_{\pi}[1]\right)_{ij}^{-1}M_{\nu;jk}[1]\left(M_{\pi}[1]\right)_{kl}^{-1}Q_l(x)}
     {Q_i(x)\left(M_{\pi}[1]\right)_{ij}^{-1}M_{\mu;jk}[1]\left(M_{\pi}[1]\right)_{kl}^{-1}Q_l(x)}
     \label{RN}
\end{eqnarray}
where $\pi$ is some positive measure, e.g $\mu$.
If we formally replace Hermitian matrix $\left(M_{\pi}[1]\right)^{-1}$ by non-Hermitian matrix 
$\left(M_{\mu}[1]\right)^{-1/2}\left(M_{\nu}[1]\right)^{-1/2}$ and its transpose
then we receive original expression (\ref{bsim}).
The (\ref{RN}) Radon--Nikodym derivative estimator can be used for interpolation
of arbitrary function $f$. Just put $\pi=\mu$ and $d\nu=f d\mu$.
See Appendix \ref{runge} as an example of Runge oscillations suppression.
The expression (\ref{RN}) in this special case $\pi=\mu$ and $d\nu=f d\mu$
is plain Nevai operator\cite{nevai} 
\begin{eqnarray}
G f&=&\frac{\int K^2(x,t) f(t) d\mu(t)}{ K(x,x)}
\label{nevai_oper}
\end{eqnarray}
That can be easily estimated numerically (see the code from com.polytechnik.utils. NevaiOperator. getNevaiOperator) as a ratio or two polynomials of order $2n$:
\begin{eqnarray}
  (G f)(x)&=&\frac{Q_i(x)\left(M_{\mu}[1]\right)_{ij}^{-1}M_{\mu;jk}[f]\left(M_{\mu}[1]\right)_{kl}^{-1}Q_l(x)}
  {Q_i(x)\left(M_{\mu}[1]\right)_{ij}^{-1}Q_j(x)}
\end{eqnarray}
This Nevai operator matches exactly the simplistic form ($\pi=\mu$) of Radon--Nikodym derivative in function approximation
like in (\ref{RNsimple}). The Radon--Nikodym derivatives approach
is based on matrices, not vectors as least square approximation is.
This matrix approach can be also effectively applied to average calculation,
see Appendix \ref{spur} for an example of two stocks price covariance calculation.

\subsection{\label{exeflow}Example of executed orders flow $I$}
In subsection \ref{radinnyk} we provided a theory
allowing to numerically calculate the executed orders flow.
To show this theory practical value let us apply it to calculation of
executed order flow $I$ for stock AAPL
on September, 20, 2012. All the charts we present will be for this
specific day. In our analysis we actually analyzed about 4 years period.
Optimized ITCH parser along with recurrent calculation of the
moments (given the moments on interval $[-\infty,-\tau]$ (old moments)
the new moments on interval $[-\infty,0]$ (t=0 is ``now'')
can be calculated using old moments and performing
timeserie scanning only on $[-\tau,0]$ interval). 
Such optimization allows to run entire trading day analysis
for hundred of stocks
in less than 15 minutes. But this massive data analysis
is not the point of the paper. This would be important
were we building some statistical arbitrage model.
But for dynamic model - single day is enough to demonstrate
the key elements of the theory. The September, 20, 2012
was chosen for simple reason that it had bear market before 10:00
and bull market with high volatility after 10:00.
Such market behavior almost always lead to severe losses
by automated trading machines, so this day is a good one for testing.

\begin{figure}
\includegraphics[width=18cm]{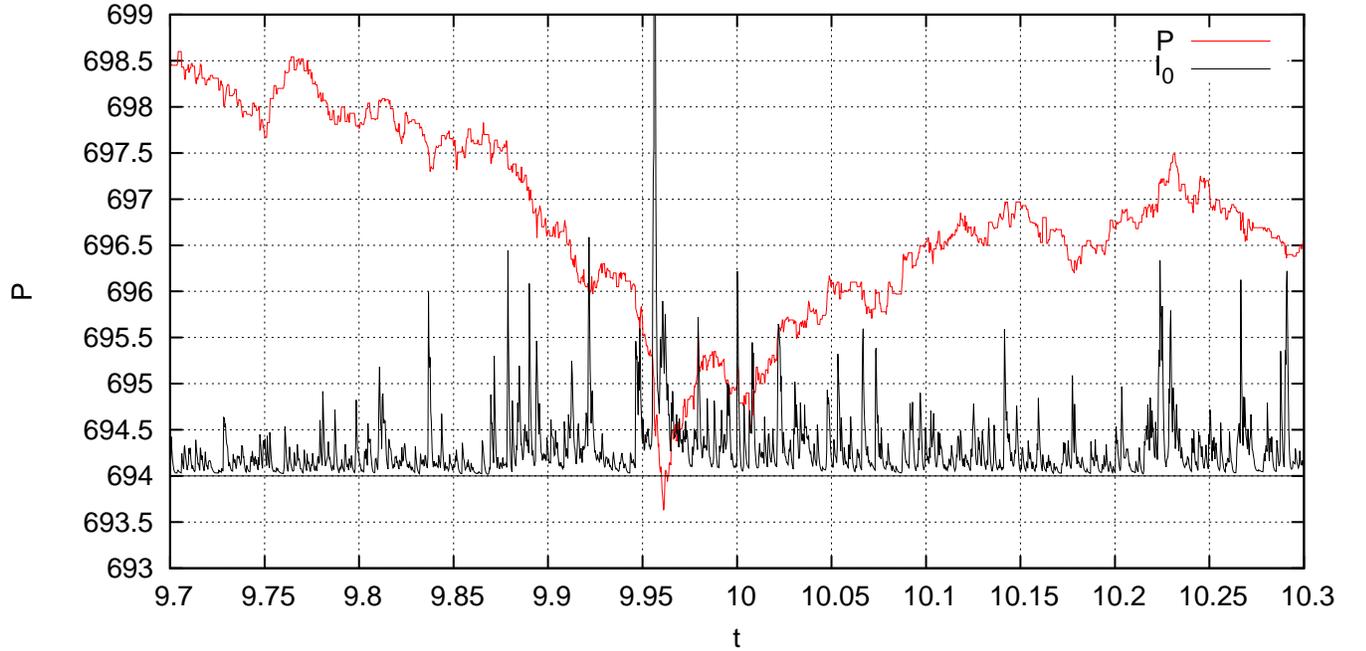}
\caption{\label{fig:I}
  The AAPL stock price on September, 20, 2012 round 10am.
  The time on x axis is in decimal fraction of an hour, e.g. 9.75 mean 9:45am.
  Red line AAPL stock price. Black line - execution order flow $I_0$ (in arbitrary units shifted to fit the chart) at interval edge(time=``now'')
  calculated in Shifted Legendre basis with $n=6$ and $\tau$=128sec.
}
\end{figure}

On Fig. \ref{fig:I} we present execution flow $I_0$ ($I$ at $t=0$)
calculated in Shifted Legendre basis as $x=\exp(t/\tau)$,
$d\mu=\exp(t/\tau)dt$ and $d\nu=\exp(t/\tau)dv$, then 
$I$ from (\ref{I}) can be estimated as Radon--Nikodym derivative $d\nu/d\mu$
calculated at $t=0$.
On Fig. \ref{fig:I} the $I_0$ (scaled to fit the chart)
show large fluctuations, with alternating periods of
low and high trading activity. High trading activity
events exhibit singular type of behavior in $I$,
that manifest itself in price as a peculiarity, not as singularity.
This allows us to suggest that in market dynamics executed trades flow
is primary and price changes are secondary.

The minimal calculable time scale of $I$ spikes can be estimated
as $\tau/(n+1)$ for Shifted Legendre basis and as $\tau$ for Laguerre basis.
If we accept the hypothesis that fluctuations in $I$ cause market dynamics
then we can estimate time scale on which automated trading machine can potentially work.
The idea is to have large fluctuations $\triangle v/\triangle t$
(fluctuations can be many orders of magnitude, but for our way to calculate
Radon--Nikodym derivatives they are limited to minimal time scale).
If minimal time scale is set large enough (about an hour)
volume fluctuations on this scale become small and such small
fluctuations of orders flow cannot be the source of predictable price movement.
In the same time too small time scale provide little liquidity
and only companies with very advanced infrastructure
can potentially take advantage of such small time scales.
This make us to conclude that workable time scales
are bounded at low values - by insufficient liquidity available
and at high values - by low $I$ fluctuations.

One can extract some additional important facts
from this chart, but the main question with $I_0$ is:
What is the scale the $I$ should be compared to to tell that
we have liquidity excess ($I>I_{IH}$) or liquidity deficit ($I<I_{IL}$).
Any values calculated from fixed time scale (e.g. $<I>$, which used $\tau$
as time scale) cannot provide workable values for $I_{IH}$ and $I_{IL}$.
The next section is dedicated to this problem.

\subsection{\label{EV}Generalized Radon--Nikodym derivatives and Generalized Eigenvectors problem}

The Eq. (\ref{RN}) can be rewritten in the form
\begin{eqnarray}
  \psi(x)&=&\mathbf{Q}(x)\left(M_{\pi}[1]\right)^{-1}\mathbf{Q}(x_0) \label{psix0} \\
  \frac{d\nu}{d\mu}(x_0)&=&\frac{<\psi|\psi>_{\nu}}{<\psi|\psi>_{\mu}} 
  \label{vRNv}
\end{eqnarray}
and for simplest case $\pi=\mu$
\begin{eqnarray}
  \psi_0(x)&=&\frac{\mathbf{Q}(x)\left(M_{\mu}[1]\right)^{-1}\mathbf{Q}(x_0)}
  {\sqrt{\mathbf{Q}(x_0)\left(M_{\mu}[1]\right)^{-1}\mathbf{Q}(x_0)}}
  \label{psix0norm} \\
  \frac{d\nu}{d\mu}(x_0)&=&\frac{<\psi_0|\psi_0>_{\nu}}{<\psi_0|\psi_0>_{\mu}} 
\end{eqnarray}
where the (\ref{psix0}) is a ``wavefunction'' localized at $x=x_0$ and
(\ref{vRNv}) is the value of Radon--Nikodym derivative at $x=x_0$.
Let us remove the localization restriction (\ref{psix0}),
then the
\begin{eqnarray}
  \frac{<\psi|\psi>_{\nu}}{<\psi|\psi>_{\mu}}=\lambda \\
  M_{\nu}[1]|\psi^{(j)}>&=&\lambda^{(j)} M_{\mu}[1]|\psi^{(j)}> \label{geq}\\
  <\psi^{(j)}|\psi^{(j)}>_{\mu}&=&<\psi^{(j)}|M_{\mu}[1]|\psi^{(j)}>=1
\end{eqnarray}
can be considered as generalized eigenvalues problem with scalar product
$<a|b>=<a|M_{\mu}[1]|b>$. The upper index $(j)$ numerate eigenvalues
and eigenvectors. If matrix $M_{\mu}[1]$ is positive, (e.g. $d\mu=\omega(t)dt$ with $\omega(t)>0$)
then (\ref{geq}) has exactly $\dim M=n+1$ real eigenvalues $\lambda^{(j)}$
and corresponding to them eigenvectors $|\psi^{(j)}>$. This problem is invariant
to basis transform. A good basis selection (e.g. (\ref{muflaguerre}) or (\ref{muslegendre}) )
make matrix $M_{\mu}[1]$ diagonal and the problem (\ref{geq}) is trivially reduced to a regular
eigenvalues problem. In general case generalized eigenvector problem
is not any more problematic, than regular eigenvalues problem and can be solved numerically using standard, e.g. LAPACK\cite{lapack} routines dsygv,  dsygvd and similar.

The problem (\ref{geq}) is much more generic than its ``localized'' Radon--Nikodym version (\ref{RN}).
Trivial usage of (\ref{geq}) is to find minimal/maximal value
of Radon--Nikodym derivative (or a function, for this just put $d\nu=f(x) d\mu$),
this will be the minimal/maximal eigenvalue $\lambda$. (Note,
that the eigenfunction, corresponding to minimal/maximal eigenvalue
has very noticeable topological properties, such as:
1. If highest order polynomial coefficient of eigenfunction $\psi(x)$ 
is non zero (if it is zero,
then it can be varied to some infinitesimal value) then the  $\psi(x)$ (a polynomial of $n$-th order) has exactly $n$
simple real distinct roots (but not necessary on the support of $d\mu$ or $d\nu$).
This property does not hold for $\psi(x)$ corresponding to other than minimal/maximal eigenvalue.
2. The measure $d\theta=d\nu-\lambda_{min}d\mu$ (or similarly for maximal eigenvalue take
$d\theta=\lambda_{max}d\mu-d\nu$) generate $n+1$ orthogonal polynomials, the last one the $n$-th order polynomial equal (within a constant)
exactly to $\psi(x)$, corresponding to $\lambda_{min}$, and has the norm with measure $d\theta$ exactly equal to zero $<\psi|\psi>_{\theta}=0$
A Gaussian quadrature can be build on this measure $d\theta$,
all nodes are located
at $\psi(x)$ roots and all weights are positive. We expect to put more study of this
interesting topic separately.)

\begin{figure}
\includegraphics[width=18cm]{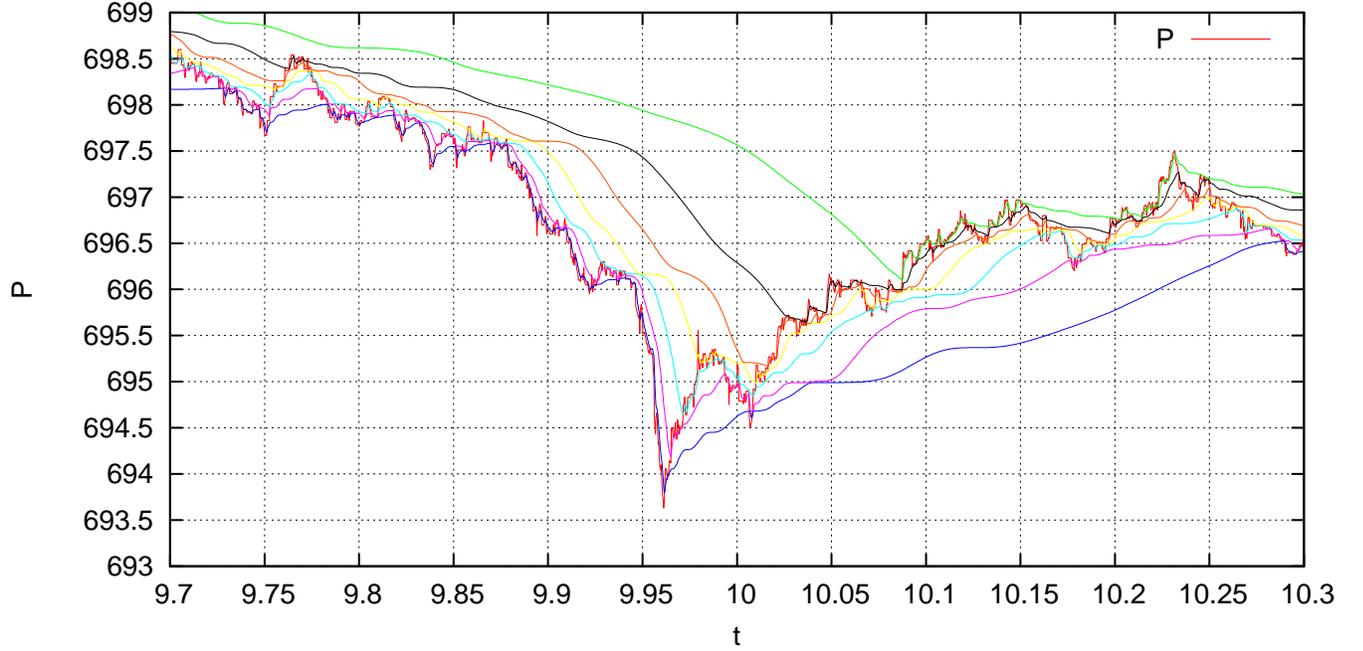}
\caption{\label{fig:qP}
   The AAPL stock price on September, 20, 2012 around 10am.
   The time on x axis is in decimal fraction of an hour, e.g. 9.75 mean 9:45am.
   Red line AAPL stock price. Other lines - eigenvalues of (\protect\ref{geq})
   with $d\nu=P(t)d\mu$,
   calculated in Shifted Legendre basis with $n=6$ (seven eigenvalues: $0..6$) and $\tau$=128sec.
}
\end{figure}

But before we go this direction, let us
show some simple illustrative example,
when $d\nu=P(t)d\mu$, where $P$ is asset price.
Then all eigenvalues are just the prices near which the asset
was traded the most. In Price basis (\ref{pspace})
the eigenvalues are the nodes of Gaussian quadrature
built on measure (\ref{mupspace}).
In Laguerre and Shifted Legendre basis
the result is very similar, but does not have a meaning
of quadrature nodes (it is now related to $M_{\mu}[P]$ matrix spectrum).
In case $n=0$ there is a single eigenvalue,
which is equal exactly to moving average with the measure $d\mu$.
So this technique can be considered as moving average generalization.
Putting price into (\ref{geq}) does not provide one with any information
about the future. The Fig. \ref{fig:qP} serve just as an illustration
of generalized eigenvalues technique.

\subsection{\label{exthr}Example of thresholds calculation}
The $\psi$ from Eq. (\ref{psix0norm}) is a state localized at $x_0$.
Consider $x_0$ to be interval end ($x_0=0$ for Laguerre basis (\ref{lagurrerange}) and $x_0=1$ for shifted Legendre basis (\ref{shiftedlegendrerange})).
All functions $\psi(x)$ orthogonal to  (\ref{psix0norm}) with respect to measure $d\mu$ have $\psi(x_0)=0$. Then we can write generalized eigenvalues equation (\ref{geq}) with a ``boundary condition'':
\begin{eqnarray}
  \frac{<\psi|\psi>_{\nu}}{<\psi|\psi>_{\mu}}=\lambda \\
  M_{\nu}[1]|\psi^{(j)}>&=&\lambda^{(j)} M_{\mu}[1]|\psi^{(j)}> \label{geq0}\\
  <\psi^{(j)}|\psi^{(j)}>_{\mu}&=&<\psi^{(j)}|M_{\mu}[1]|\psi^{(j)}>=1 \\
  \psi(x_0)&=&0 \label{psi0boundary}
\end{eqnarray}
The boundary condition (\ref{psi0boundary}) can be removed by introducing
two measures $d\widetilde{\mu}=(x-x_0)^2d\mu$ and $d\widetilde{\nu}=(x-x_0)^2d\nu$, then
\begin{eqnarray}
  \frac{<\phi|\phi>_{\widetilde{\nu}}}{<\phi|\phi>_{\widetilde{\mu}}}=\lambda \\
  M_{\widetilde{\nu}}[1]|\phi^{(j)}>&=&\lambda^{(j)} M_{\widetilde{\mu}}[1]|\phi^{(j)} \label{geq0phi}\\
  <\phi^{(j)}|\phi^{(j)}>_{\widetilde{\mu}}&=&1 \\
  \psi(x)&=&(x-x_0)\phi(x) \label{psiphi}
\end{eqnarray}
we receive regular generalized eigenvalues problem and $\psi$ from (\ref{psiphi})
obey the required boundary condition (\ref{psi0boundary}).
Any solution $\psi$ of (\ref{geq0phi}) is orthogonal to (\ref{psix0norm}) because
it is equal to $0$ at $x_0$, e.g. it carry no information about ``now'',
only about ``the past''. The minimal/maximal eigenvalues of (\ref{geq0phi})
 $I_{IL}$ and $I_{IH}$ are the thresholds we were looking for\footnote{
  There are other ways to estimate thresholds. One
  can use e.g. Radau-like measures \hbox{$d\widetilde{\mu}=(x-x_0)d\mu$}
  (or \hbox{$d\widetilde{\mu}=(x_0-x)d\mu$} depending on $x_0$ position)
    or, another option, drop boundary condition (\protect\ref{psi0boundary}) altogether
    and estimate whether $I$ is ``low'' or ``high'' as
    closeness of $|\psi_{\{IL,IH\}}>$ to $|\psi_0>$, localized at $x_0$ (the one from (\ref{psix0norm})) as  $<\psi_{\{IL,IH\}}|\psi_0>_{\mu}^2$.
    We will discuss this approach separately.}.

\begin{figure}
\includegraphics[width=18cm]{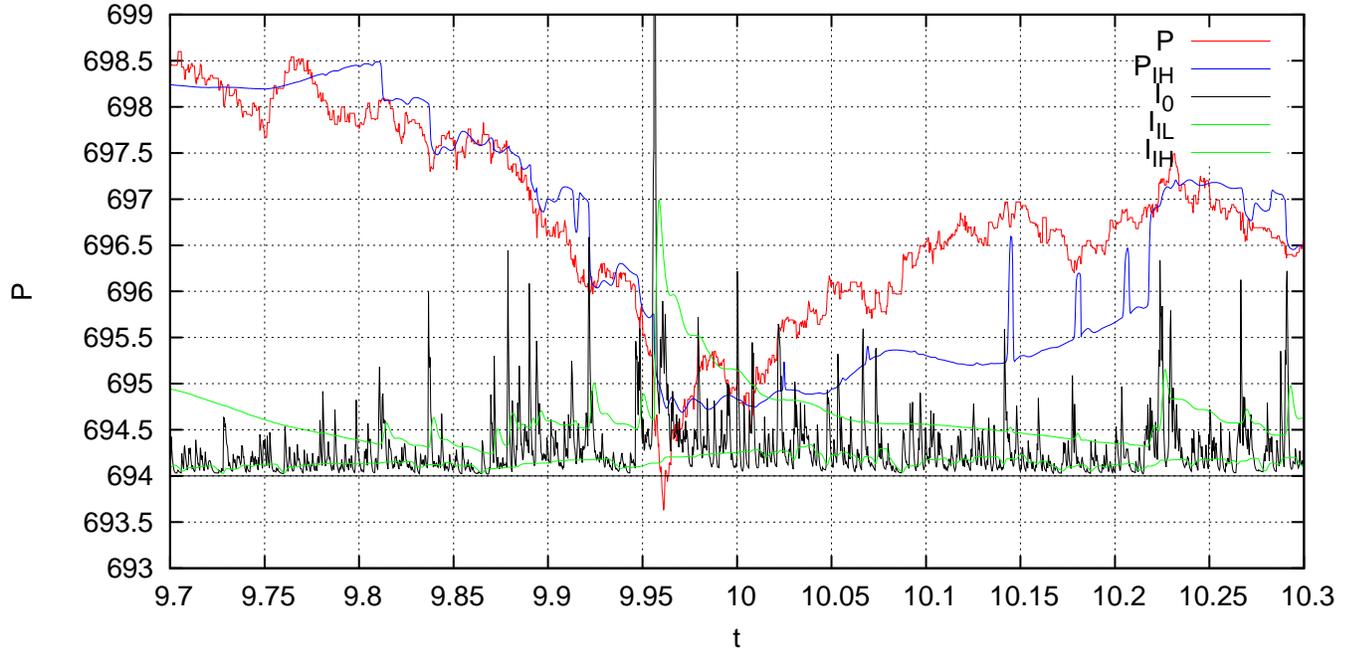}
  \caption{\label{fig:Ith}
    Same chart as Fig. \protect\ref{fig:I} with addition $I_{IL}$ and $I_{IH}$
    thresholds and price (corresponding to $\psi_{IH}$) are calculated.
    The $I_{IL}$ and $I_{IH}$ are minimal and maximal eigenvalues
    of (\ref{geq0phi}). Because of the boundary condition the corresponding matrix has the dimension
    one less the original matrix dimension.
}
\end{figure}

On the Fig. \ref{fig:Ith} we present execution rate $I_0$ and thresholds
$I_{IL}$ (``low'', minimal eigenvalue) and $I_{IH}$ (``high'', maximal eigenvalue) calculated only from ``the past''.
One can observe two highly distinctive behavior at $I_0<I_{IL}$ (liquidity deficit)
and $I_0>I_{IH}$ (liquidity excess).
It is important to note that the time scale corresponding
to $I_{IL}$ and $I_{IH}$ is not fixed, but selected automatically
from the time scales available in matrix $M_{\mu}[I]$
(for a matrix of dimension $d$ the $2d-1$ time scales are used).
The events when $I_0>I_{IH}$ are rather seldom,
and as we would show later they are exactly the
events portfolio position to be closed.
The events $I_0<I_{IL}$ are much more common
and as we would show later they are exactly the
events portfolio positions to be opened.
The price $P_{IH}$ (blue curve)
is the price corresponding to $\psi_{IH}=<\psi_{IH}|pI|\psi_{IH}>_{\mu}/<\psi_{IH}|I|\psi_{IH}>_{\mu}$ (as a very crude direction estimator
a difference between last price and $P_{IH}$ can be used).
Similarly to $I=dv/dt$ same theory can be applied to
calculation of $dp/dt$ at $t=0$ and
corresponding thresholds for $dp/dt$.
The problem with $dp/dt$ is the contribution to $dp/dt$ at $t=0$
is so large that it exceeds the thresholds calculated on past data
most of the time.
This again manifest our statement that price alone carry no information
about dynamic.

\begin{figure}
\includegraphics[width=18cm]{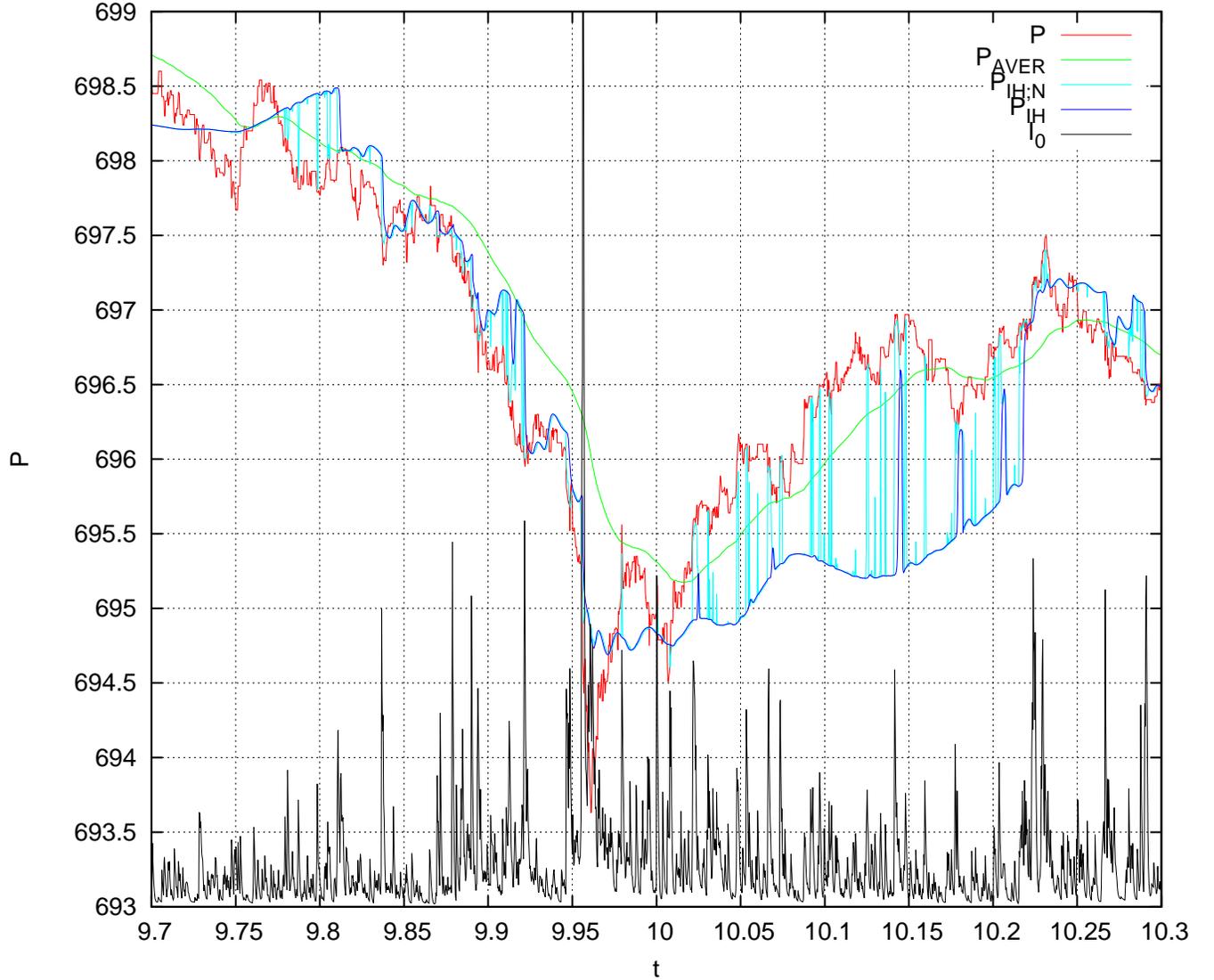}
\caption{\label{fig:qPaver}
  The AAPL stock on September, 20, 2012 around 10am.
  Price $P$, $P_{IH}$, $P_{IH;N}$ and $P_{AVER}$ are presented.
}
\end{figure}

On Fig.\ref{fig:qPaver} we present one more 
chart  to show prices behavior  at various $|\psi>$ states.
For $n=6$ and $\tau=128 sec$ the
$P_{IH;N}$ calculated from generalized eigenvalues problem
without using boundary condition $\psi(x_0)=0$ (matrix dimension is $n+1$),
$P_{IH}$ calculated from generalized eigenvalues problem (\ref{geq0phi})
with boundary condition $\psi(x_0)=0$
(the original matrix dimension is $n+1$, but boundary condition (\ref{psiphi})  reduce it by 1 to use the same moments),
and exponential moving average $P_{AVER}$.
What one can very clear see is while $P_{AVER}$
is always delayed from the stock price by a fixed time $\tau$ the delay for
$P_{IH}$ is variable and depend on localization of $\psi_{IH}$.
This is very important for market trending identification:
one do not need to wait time $\tau$ to identify trend change.
The $P_{IH;N}$, is a solution of similar eigenproblem,
but without  boundary condition $\psi(x_0)=0$.
When $I_0$ ($I$ at $x_0$ or $t=0$, i.e. ``now'') is high
then the solutions for the problems with and without boundary
condition differ(one has $\psi(x_0)=0$, another one is localized at $x_0$)
and corresponding prices (dark and light blue) also differ significantly
(this difference, calculated on
liquidity excess events can also serve as a crude
estimator of market trending direction).
When $I_0$ is low then  $P_{IH}$  and $P_{IH;N}$ are almost the same
because corresponding states $|\psi>$ are not localized near $x_0$.
What is the most important --  the states corresponding to
maximal\footnote{
  The states corresponding to minimal $I$ poses similar
  properties, but  as we noted in Section \protect\ref{EV}
  the $\psi_{IL}(x)$ roots
  are simple real distinct (but not necessary on the support of the measure).
  Because $I$ is the lowest on this state, the high values of $I$
  should be localized near the $\psi_{IL}(x)$ roots).
  In most situations
  the results obtained near the $\psi_{IL}(x)$ roots
  are very similar to the calculations above on $|\psi_{IH}>$ state.
} $I$ select the timescale automatically
among the ones available in matrix $M_{\mu}[I]$,
what is drastically different from moving averages,
which has only a fixed time scale $\tau$.

\subsection{\label{PnL}P\&L operator and trading strategy}
Before we go further
we would like to emphasize the importance of variables selection.
As we discussed earlier the price fluctuations
are small (below few percent) and only reflect 
liquidity fluctuations.
Nevertheless most traders and Automated Trading Machines
focus on price prediction. From our opinion
prices cannot be predicted on real markets.
But if you look deeper, a trader is not actually interested
in prices, what actually of his interest is the P\&L.
From our point of view the P\&L, not price, should be a value to predict.
Let us define the position change $dS$ - the amount of shares
bought ($dS>0$) or sold ($dS<0$) during time interval $dt$.
Then the P\&L can be written in the form:
\begin{eqnarray}
  \mathrm{P\&L}&=-&\int p dS \label{PL} \\
  0&=&\int dS \label{PLconstrain}
\end{eqnarray}
The constrain (\ref{PLconstrain}) means the
total asset position should be zero in the
beginning and in the end of trading period.
Integrating (\ref{PL}) by parts one can obtain
a P\&L expressed via price changes $dp$:
\begin{eqnarray}
  \mathrm{P\&L}&=&\int S(t) dp \label{PLdp} \\
  0&=&S(t_{start})=S(t_{end}) \label{PLconstraindp}
\end{eqnarray}
where constrain (\ref{PLconstraindp}) explicitly
indicate that the position $S(t)$ should be 0 in the beginning
and in the end of trading interval.

Now the problem can be formulated in the following way:
find the position function $S(t)$ providing positive P\&L.
There is a trivial solution: $S=dp/dt$ to put to (\ref{PLdp})
or $dS=dt d^2p/dt^2$ to put to (\ref{PL}). This means
that position increment $dS$ should behave as second derivative
of price. This sounds trivial (if you know future price change you can make
money), but it is actually not. The very important
is the symmetry of position increment. Position increment should have
the symmetry of second derivative of price
(first derivative is good only for entire position, not position increment.
An Automated Trading Machines trading in position increment
but using a variable with a symmetry of first price derivative
cannot give a success).

Let us give some other trivial, but nevertheless useful examples
of  position function $S(t)$  providing positive P\&L.

Assume we have sufficient liquidity to buy shares in any time moment
and trade a single share in just two moments (sell/buy or buy/sell) of unit length.
Then we can take position increment in the form
\begin{eqnarray}
  dS&=&\psi^2_{buy}(x)d\mu - \psi^2_{sell}(x)d\mu \label{PLxample} \\
  1&=&<\psi^2>_{\mu} \label{PLnorm}
\end{eqnarray}
for normalized $\psi$ (\ref{PLnorm})
the condition (\ref{PLconstrain}) satisfies automatically
and the problem (\ref{PL}) is reduced to the following
generalized eigenvalues problem:

\begin{eqnarray}
  &-&\left[<\psi_{buy}|M_{\mu}[p]|\psi_{buy}>-
    <\psi_{sell}|M_{\mu}[p]|\psi_{sell}>\right]= \nonumber \\
  &&\mathrm{P\&L}
  \left[<\psi_{buy}|M_{\mu}[1]|\psi_{buy}>
    -<\psi_{sell}|M_{\mu}[1]|\psi_{sell}>\right]
  \label{PLgevexample} \\
  \mathrm{P\&L} &\to& \max
\end{eqnarray}
the solution for maximal $\mathrm{P\&L}$ in (\ref{PLgevexample})
is rather trivial. Solve generalized eigenvalues
problem $M_{\mu}[p]|\psi>=\lambda M_{\mu}[1]|\psi>$
then take $\psi_{buy}$ as $\psi$ corresponding to minimal $\lambda$
and take $\psi_{sell}$ as $\psi$ corresponding to maximal $\lambda$,
then $\mathrm{P\&L}=\lambda_{max}-\lambda_{min}$.
The answer is trivial buy low ($p=\lambda_{min}$) and sell high ($p=\lambda_{max}$)
and not practical (as we stated earlier price carry no information
about future price change)
but nevertheless very useful: it indicates the power of the technique:
$\mathrm{P\&L}$ optimization problem is reduced to matrix spectrum analysis.

Another trivial example: hold some fixed average position.
\begin{eqnarray}
  S&=&\psi^2(x) \label{PLxampleS} \\
  1&=&<\psi^2>_{\mu} \label{PLnormS} \\
  0&=&\psi(t=-\infty)=\psi(t=0) \label{PLnormbcS}
\end{eqnarray}
The (\ref{PLnormS}) set average position held and
boundary condition (\ref{PLnormbcS}) require no position to remain
outside of trading interval.
Then using (\ref{PLdp}) we receive:
\begin{eqnarray}
  &&<\psi|M_{\mu}[dp/dt]|\psi>=\mathrm{P\&L}
  <\psi|M[1]|\psi>  \label{PLgevexampledp} \\
  \mathrm{P\&L} &\to& \max
\end{eqnarray}
which has a simple solution:
Solve generalized eigenvalues
problem $M_{\mu}[dp/dt]|\psi>=\lambda M_{\mu}[1]|\psi>$,
find $\lambda_{min}$ and $\lambda_{max}$, select
the one with maximal absolute value, the corresponding $\psi$ is the answer.
This answer is also trivial if market go up ($dp/dt>0$) hold long position,
if market go down ($dp/dt<0$) hold short position.
Again, the example is not practical,
it just indicates how $\mathrm{P\&L}$ optimization problem is reduced to matrix spectrum analysis.
One more note about P\&L is that it is typically calculated
on cash basis (require no shares held outside trading interval, then calculate
P\&L cash difference), but for some trading strategies asset-based
definition can be more useful (require no cash held outside trading interval,
then calculate P\&L as shares number difference).

\section{\label{dynamics}Dynamics}

\subsection{\label{obser}Observable and Unobservable variables}
\begin{figure}
\includegraphics[width=14cm]{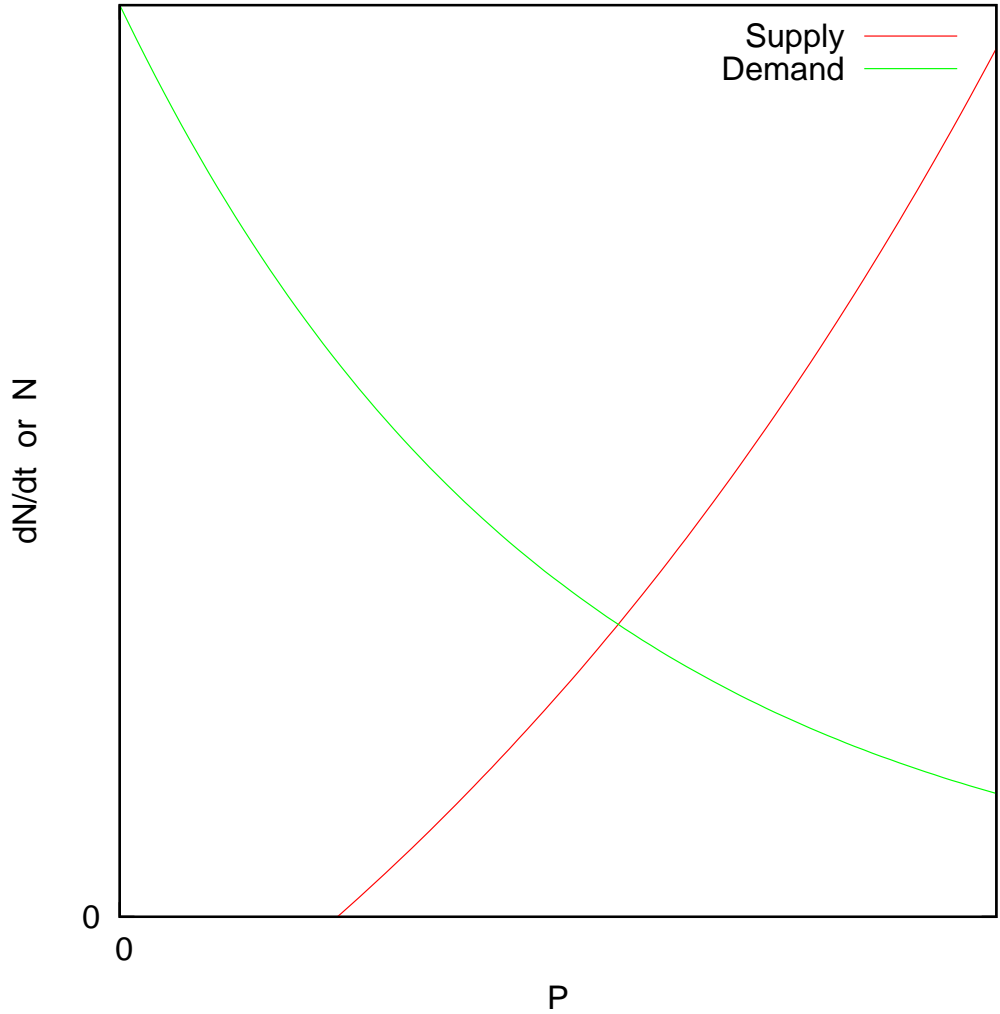}
\caption{\label{fig:qSD}
  Schematic plot of supply and demand as a function of price.
}
\end{figure}

What variables can be potentially used for market dynamics?
We already worked with such variables
as price $p$ and  executed orders flow $I=dv/dt$.
They are real, they are reported on execution tape by exchanges.
There are other variables, which are slightly more difficult to observe, e.g.
spread, order type (buy/sell), time the order type was put to order book,
orders distribution in order book, etc.
And there are other, ``virtual'' variables, such as supply or demand.
A schematic supply---demand chart is presented on Fig. \ref{fig:qSD}.
We will treat supply and demand  as flow (number or units in unit time $dN/dt$),
not as total number of units. If some supply-demand chart
is stationary and has a form similar to Fig. \ref{fig:qSD} it is
clear that only the price corresponding $dN_{buy}/dt=dN_{sell}/dt=I$
is the stationary solution and   execution take place
only at this equilibrium price. When, for any reason, execution
take place at price, different from the equilibrium the supply-demand disbalance 
formally give orders accumulation with time.
This accumulation actually never happen
in practice (either orders flow stops or price changes),
but the accumulation can be formally considered as an increase in
limit order execution time.
But limit order execution time
is actually known, this is the time the order spent in the order book
before execution.
The product of signed $I$ by time the limit order spent in the order book
before execution can serve as a supply-demand estimator.
We are going to discuss observable supply-demand estimators
in a separate publication, and touch here only fundamental
properties having the goal to transform supply=demand condition
to the one expressed only in terms of observable variables.

The question: what can we tell about supply and demand curves
at prices different from equilibrium one.
The answer is: nothing. The orders flow at prices
off current is not measurable and, we would tell even stronger,
actually do not exist at any price except currently executed
(unexecuted order book orders flow
is not a supply/demand, this is just manipulations and traders pipe dreams).

Stationary chart like Fig. \ref{fig:qSD}
or even non-stationary supply-demand dependencies are conceptually incorrect
in equity trading, because it operates with values, which cannot be measured
or even estimated. A theory can work
with unobservable concept, e.g. our theory, same as quantum mechanics,
operate with $\psi(x)$, but only $\psi^2(x)$ enter into measurable values.
The supply=demand classical approach
can be replaced by the one working only with observable variables:
\begin{eqnarray}
  I(p)&\to&\max \label{Ip}
\end{eqnarray}
The (\ref{Ip}) means : ``the price tend to the value,
maximizing future $I(p)$''.
The stationary theory on Fig. \ref{fig:qSD} is equivalent to Eq. (\ref{Ip}),
reverse is not true and the (\ref{Ip}) is much more generic and can be applied
to securities trading dynamics.
Critically important that (\ref{Ip}) operates only with observable
variables (observable postfactum, the $I$  we were calculating in Section \ref{kin}
is calculated on past(already observed) values, but even this is much better
than supply=demand classical theory where the values of supply and demand
cannot be measured even postfactum.

\subsection{\label{volatility}Volatility}
Price volatility is a very old concept, and ``reverse-to-the mean''
type of theories is actually equivalent to:
price tend to the value, at which 
volatility (measured as standard deviation calculated on past sample)
is minimal.
\begin{eqnarray}
  \mathrm{Volatility}&=&\left<(p-P_{AVER})^2\right> \\
  \mathrm{Volatility}&\to&\min \\
  P_{AVER}&=&<p>/<1> \label{Peq}
\end{eqnarray}
while this type of strategy would never work in practice (see Section \ref{speculations}
for description of the reasons), there are critically important questions:
What volatility actually is? Does ``true'' volatility correspond more or less
to price fluctuation or to $I$ fluctuation?
Is volatility a concept of the same nature as $I$
or they are completely different concepts?
Looking at charts we see that price volatility is typically large at large $I$, but this may be like kinetic to potential energy transform in mechanics.

The other definitions of volatility can be introduced
as price fluctuations, e.g.
$\mathrm{Volatility}=\left<(dp/dt)^2\right>$ , the problem
with this definition is that it diverges at small time scales.
(One derivative is compensated by the integral, and another one
is translated to measure support boundary, what lead to
expression divergence at small time scales.)
The $M_{\mu}[(dp/dt)^2]$ matrix cannot be directly calculated
from price timeserie sample, and the formal expansion in a style of Appendix \ref{spur}
 $M_{\mu}[(dp/dt)^2]=M_{\mu}[dp/dt]G_{\mu}^{-1}M_{\mu}[dp/dt]$
is not a good one because it introduces basis minimal scale into the expansion.

Let us give alternative volatility definition:
\begin{eqnarray}
  \mathrm{Volatility}&=&\left<|dp/dt|\right> \label{dpVolatility}
\end{eqnarray}
This definition uses first derivative, so
all the moments can be directly calculated from price timeserie sample,
as $\int Q_k(x)\omega(x) |dp|$, this expression is essentially the same as
$dp/dt$ moments, but absolute value of price change
should be used in the sum corresponding to the integral.
Technically this calculation is almost the same as $dv/dt$ moments calculation,
with the difference that ``trading events'' occur in the points of price
change and the ``trading volume'' is absolute value of price change.
\begin{figure}
\includegraphics[width=18cm]{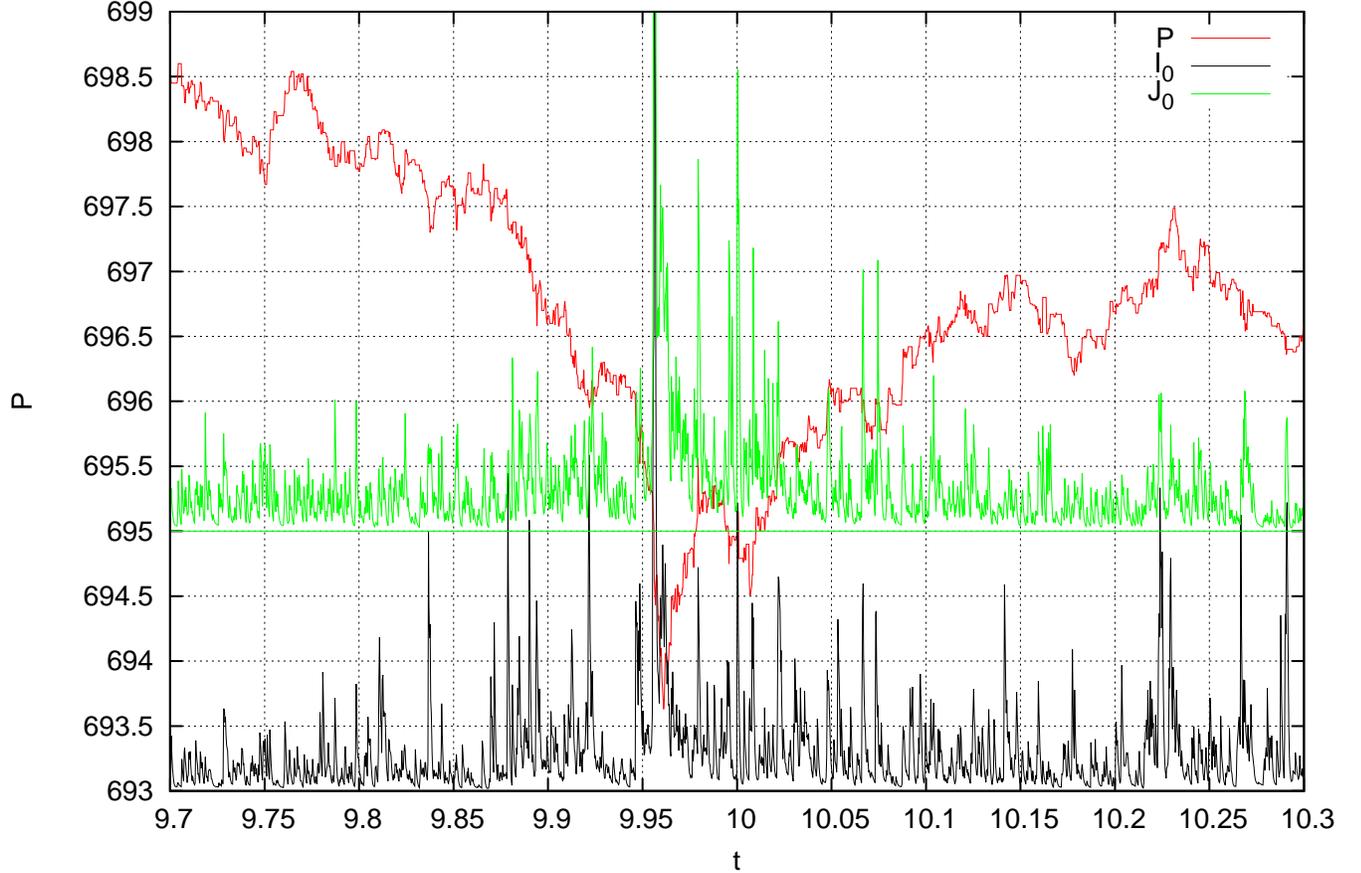}
  \caption{\label{fig:Iadp}
    A chart similar to the Fig. \protect\ref{fig:I},
    but comparing real execution flow $I=dv/dt$ (black line) and
    artificial one $J$ (green line) calculated from $|dp|/dt$.
}
\end{figure}
On the Fig. \ref{fig:Iadp} we present real execution flow $I=dv/dt$ (black line) and
artificial one $J=|dp|/dt$ (green line). They are very similar in nature.
This probably means that supply-demand and price volatility are the entities of the same nature, at least
for equity trading.
When trading volume is unavailable the $|dp|$ can be used
as a substitute of $dv$.

During our attempt to build dynamic equation
we spent substantial effort in an attempt to define
Lagrange functional $L$ and then build action $\cal S$ like in 
other dynamic theories:
\begin{eqnarray}
  L&=& \frac{m}{2}\mathrm{Volatility}-I \label{Ldyn}\\
  {\cal S}&=& \int L dt  \label{action} \\
  {\cal S}&\to& \min \\  
  \delta {\cal S}&\to& 0
\end{eqnarray}
This approach is very attractive: it requires to minimize price volatility
(like in ``reverse-to-mean'' type of theories)
and to maximize execution flow $I$ (like in supply-demand theories),
but possibly fruitless.
We spent substantial time pursuing this
route using various volatility models and constrains on action variation
with no improvement compared to using just $I$
and only supply-demand functional (\ref{Ip}).
This means that the ``effective mass'' $m$ in (\ref{Ldyn})
is close to $0$, at least for equity trading. At this point
we do not have an answer
to the fundamental question about price volatility role,
and whether other therms, similar in their nature to volatility,
should be put to Lagrange functional (\ref{Ldyn})
along with supply-demand term $I=dv/dt$.
In all the calculations below we would assume $m=0$.
Note that in stationary case on Fig. \ref{fig:qSD} these volatility--like
terms play no effect, they play role only in dynamic
situation, when price change, so our approach can be
considered as a ``quasistationary approximation''.
We can give another reason why price volatility (but not the terms like $\left(d\psi/dt\right)^2$, that lead to Schr\"{o}dinger --- like equation,
which was also tried without much success) should not enter
the dynamic equation: as we discussed earlier, price
fluctuations are secondary to liquidity fluctuations, and
position enter/exit conditions should be calculated
without price used. Then, only on the last step,
when P\&L need to be calculated the price should be used to calculate the direction.

\subsection{\label{dynP}Price corresponding to maximal $I$ on past sample}
The $P_{AVER}$ introduced in Subsection  \ref{volatility}
is calculated
as average over some time (or volume) interval (\ref{Peq}).
This price(calculated on past sample) has no any degree of freedom available
and correspond to a strategy buy below $P_{AVER}$, sell above $P_{AVER}$ thus maximize trading volume
(to have the condition (\ref{PLconstraindp}) satisfied
one have to use median, not average price, but for practical
calculations median and average are close enough).
Now, instead of trading to maximize volume
consider trading to maximize  $I$, Eq. (\ref{Ip}).
What is different, we now have $\dim M$ degrees
of freedom ($\psi$ components) available, that are selected
to have $I$ maximized.
Calculations still uses only ``charted'' past prices
(because both measures $Id\mu$ and $d\mu$ are positive),
but the time scale is now selected automatically. This is
the most critical improvement when doing a transition
from maximizing volume to maximizing $I$.
The corresponding price $P_{IH}$ (two versions
calculated with different boundary conditions
were already
presented on the Fig. \ref{fig:qPaver}), but now we
are going to perform an
 analysis it in terms of P\&L dynamics.

The problem can be formulated as to find a strategy,
maximizing the P\&L. Let us present a simple, but nevertheless practical,
trading strategy, which
exhibit all the important elements of the theory.

Input: at time $t_i$ execution with price $p(t_i)$ and trading volume $dv(t_i)$.

Continuously calculate
$I_0$, $I_{IL}$, $I_{IH}$, and $P_{IH}$ as we did in the previous section.
There is one more variable $\mathit{dir}$,
which determine the direction of position opening,
and threshold constant $\mathit{th}$, that is typically selected about 0.8-0.9.
Then apply the following heuristics:

\begin{enumerate}
\item \label{entercond} If $I_0<I_{IL}$ enter long position if $\mathit{dir}>\mathit{th}$,
enter short position if $\mathit{dir}<-\mathit{th}$,
otherwise hold no position.
\item \label{exitcond} If $I_0>I_{IH}$:
  \begin{itemize}
    \item Recalculate $\mathit{dir}$.
    First
    calculate $P_{IH}$ (see Section \ref{exthr}) with boundary condition $\psi(x_0)=0$, then build matrix $M_{\mu}[(p-P_{IH})I]$ from $M_{\mu}[p]$ and
    $M_{\mu}[pI]$ matrices calculated directly from the moments of observable samples.
    The matrix $M_{\mu}[(p-P_{IH})I]$ corresponds to P\&L matrix
    in scenario ``enter position at $\psi_{IH}$''. Note that if entering
    position take unit time, then the $I_{IH}$ is the maximal
    volume which can be accumulated in unit time on past sample.
    \item
Determine how ``exit now'' scenario is good for P\&L operator.
Solve generalized eigenvalues problem (without boundary condition $\psi(x_0)=0$)
$\frac{<\psi|M_{\mu}[(p-P_{IH})I]|\psi>}{<\psi|M_{\mu}[1]|\psi>}=\lambda_\mathrm{P\&L}$,
find $\psi_{\mathrm{P\&L} ; \min}$ and $\psi_{\mathrm{P\&L} ; \max}$, corresponding
to $\min$ and $\max$ values of $\lambda_\mathrm{P\&L}$, then
\begin{eqnarray}
  \mathit{dir}&=&<\psi_0|\psi_{\mathrm{P\&L} ; \max}>^2-<\psi_0|\psi_{\mathrm{P\&L} ; \min}>^2\label{mdir}
\end{eqnarray}
where $\psi_0$ is from Eq.(\ref{psix0norm}).
\item Remember $\mathit{dir}$ for later use on stage \ref{entercond}.

\item If $\mathit{dir}>\mathit{th}$ close long position,
  if $\mathit{dir}<-\mathit{th}$ close short position.
\end{itemize}
\end{enumerate}
Conceptually the described heuristics is similar to $p_{last}-P_{IH}$ directional trading
(all supply-demand type of theories are directional
theories), but generalized eigenvalues techniques is used
to estimate the thresholds and time scale.
Note that if one need just a prediction of $I$ -
the result is very accurate: If current $I_0$ is large ($I_0>I_{IH}$)
then future $I_0$ will be low ($I_{0}<I_{IL}$), similar
if current $I_0$ is low ($I_{0}<I_{IL}$), then future $I_0$ will be high ($I_0>I_{IH}$).
This may look trivial (alternating periods of low and high liquidity availability)
but this mean that liquidity(not price!) undergo large oscillations,
and price changes are just the consequences of large changes in liquidity.

The key element of the strategy is that it actually trades
liquidity, providing liquidity during deficit and taking it during excess.
Our HFT experiments to be discussed in details in a separate publication,
here we just put briefly only the most important qualitative observations.

Among a number of different strategies tested ---
only this one provided no eventual catastrophic  P\&L drain
(``Black Swan''\cite{taleb2010black} --like events).
The reason is simple: the strategy
of holding zero position during liquidity excess
make the system resilient to the situation when market
moves against position held, but in the same time
entering the position during liquidity deficit (when the volatility is small)
make the system collecting most of the  market movement juice. 

Our experiments (especially for other than equity markets)
show that in a situation when market direction is known by a human trader
the value of $\mathit{dir}$ can be set manually 
according to trader's view and the system
would effectively collect the P\&L on small market movements,
in the same time avoiding catastrophic  P\&L drains
on the events when market moves against position held.
\begin{figure}
\includegraphics[width=18cm]{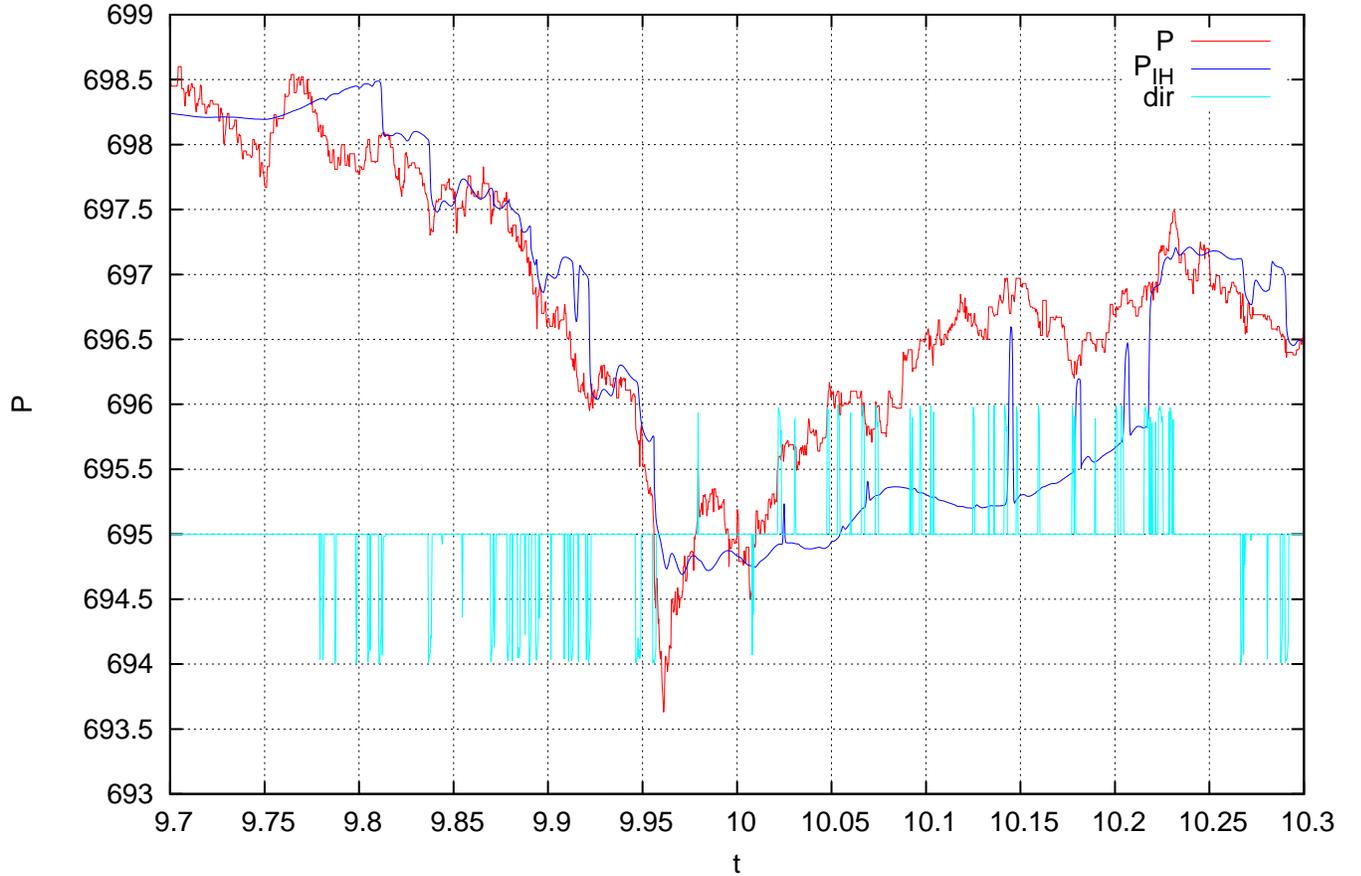}
\caption{\label{fig:Idir}
  The AAPL stock on September, 20, 2012 around 10am.
  Price $P$, $P_{IH}$, and $\mathit{dir}$ (at $I_0>I_{IH}$) are presented.
}
\end{figure}
On Fig. \ref{fig:Idir} we present calculated $\mathit{dir}$ for $I_0>I_{IH}$.
The calculated during liquidity excess the  value of $\mathit{dir}$ should be saved for
liquidity deficit ($I_0<I_{IL}$)
time moments for determination of position opening direction.
The chart shows time scale auto adjustment, what is drastically different
from $P_{AVER}$ on Fig. (\ref{fig:qPaver}), where time scale
is exactly $\tau$. The result is stable in a sense
the time scale of $\mathit{dir}$ sign change is greater than
minimal time scale available in $M_{\mu}[I]$ matrix.

Testing the strategy on real data (even paper trading, not to
mention real trading) is a complex task, because
all the fees, commissions, delays should be taken into account.
In this paper we will give qualitative description of results
obtained as a ``paper trading'' on four year period,
the detailed results to be published elsewhere.

Any attempt to use $P_{AVER}$ (corresponding to maximizing trading volume on past trades)
give losses. When used in trade
following strategy because of $\tau$ delay in trend switch identification.
There are relatively small losses on almost all days.
When using ``reverse to $P_{AVER}$'' type of strategy
most of days are profitable, but because of catastrophic P\&L drain
on relatively seldom trending days (like the one we used
in this presentation) overall P\&L is negative.
Use of $P_{IH}$ (corresponding to maximizing $I=dv/dt$ on past trades)
to determine market direction $\mathit{dir}$
and then using this direction to enter position during liquidity deficit and
closing position during liquidity excess typically give
profit on both volatility days and trending days.
Very important is that this strategy give no days with catastrophic P\&L drain.
Total number of trades per day is about few hundred for high
liquidity stocks. Average daily return vary from -1\% to 2\%
depending how execution price is modeled and exchange commissions.
Our main result is that self--adjusting time scale and liquidity(not price) based
enter/exit conditions is critically important for
a reasonable Automated Trading Machine. Our approach to market
dynamics as maximizing $I(p)$, Eq. (\ref{Ip}), (even on past trades,
what we do in this paper,
without volatility terms discussed in Section \ref{volatility}, that are not well understood), give
very promising results.
The most important is experimental evidence that there is no catastrophic P\&L drain in liquidity trading strategy.

\subsection{\label{voltrading}Volatility Trading}
In previous section \ref{dynP} we did out best to
build directional trading machine, that was
trying to predict, with some success, future price.
But price prediction is extremely difficult because,
as we stated earlier, price fluctuations are small and are secondary to
liquidity fluctuation, so our P\&L trading theory
from section \ref{PnL} was an attempt to overcome this issue.
A question arise whether  liquidity deficit
can be traded directly. If we accept experimentally
observed in section \ref{volatility} fact that
liquidity deficit is an entity of the same nature
as volatility then the answer is yes, and
liquidity deficit can be traded through some kind of derivative instruments.
Let us illustrate the approach on a simple case -- options trading.
Whatever option model is used, the key element of it is implied volatility.
Implied volatility trading strategy can be implemented
through trading some delta--neutral ``synthetic asset'',
built e.g. as long--short pairs of a call on an asset and an asset itself, call--put pairs or similar ``delta--neutral vehicles''.
Optimal implementation of such ``synthetic asset''
depends on commissions, liquidity available, exchange access, etc.
and varies from fund to fund. 
Assume we have built such delta--neutral instrument, the
price of which depend on volatility only.
How to trade it? We have the same two requirements:
1) Avoid catastrophic  P\&L drain and 2) Predict future value of volatility
(forward volatility).
Now, when trading delta--neutral strategy, this
matches exactly our theory and trading algorithm
become  this.

\begin{enumerate}
\item \label{deriventer} If for underlying asset we have $I_0<I_{IL}$
  then enter ``long volatility''
  position for ``delta--neutral'' synthetic asset.
  This enter condition means that if current execution flow
  is low - future value of it will be high, what exactly
  correspond to price dynamics from section \ref{dynP}:
  If at current price the value of $I_0$ is low -- the price
  would change to increase future $I$.
\item \label{derivexit} If for underlying asset we have $I_0>I_{IH}$
  then close existing ``long volatility''
  position for ``delta--neutral'' synthetic asset.
  At high $I_0$ future value of $I$ cannot be determined,
  it can either go down(typically) or increase even more(much more seldom,
  but just few such events sufficient to incur catastrophic P\&L drain).
  According to main concept of our P\&L trading strategy,
  one should have zero position during market uncertainty.
\end{enumerate}
The reason why this strategy is expected to be profitable
is that experiments
show that implied volatility is very much price fluctuation--dependent,
and execution flow spikes $I_0>I_{IH}$ in underlying asset
typically lead to substantial price move of it
and then implied volatility increase for ``synthetic asset''.
This strategy is a typical ``buy low volatility'', then ``sell high volatility''.
The key difference from regular case is that, instead of price volatility,
liquidity deficit is used as a proxy to forward volatility.
The described strategy never goes short volatility,
so catastrophic  P\&L drain is unlikely.
We performed the strategy testing on much more
limited data we have available to us (about 1 month of CME data) than
we did for testing directional strategy on data for
NASDAQ ITCH\cite{itchfeed} (4 years of data),
but the effect, nevertheless, clearly exist, but more testing is required
to get the final conclusion about applicability of liquidity deficit as a proxy to implied volatility.
In addition to that we want to emphasize,
that despite our theory seems to predict implied volatility much better than price direction,
actual trading implementation require the use of
``delta--neutral'' synthetic asset,
what incur substantial cost on commissions and execution,
thus actual P\&L is difficult to estimate without existing setup for
high--frequency option trading.

\section{\label{speculations}Speculations}
In this paper we presented a theory trying to describe kinematics and dynamics
of the market.
The effect is relatively weak, so it is difficult to make money directly,
but provided theory can state very clear what kind of Automated Trading Machines
CANNOT make money. In best case they will be making
little money for some time, then lose more than they made in a single event.
Specifically:
\begin{itemize}
\item Any system that uses only single asset price
  (and possibly prices of multiple assets, but this case is not completely clear)
  as input. The price is actually secondary and typically fluctuates
  few percent a day in contrast with liquidity flow, that
  fluctuates in orders of magnitude. This also allows to estimate
  maximal workable time scale: the scale on which execution flow
  fluctuates at least in an order of magnitude (in 10 times).
\item Any system that has a built-in fixed time scale (e.g. moving average type of system).
  The market has no specific time scale.
  Minimal number of time scales is 3 (the time scales of 2x2 matrix (\ref{mc}),
  typical value to make system some-kind working is 13 time scales 
  (all time scales of 7x7 matrix (\ref{mc})).
\item Any ``symmetric'' system with just two signal ``buy'' and ``sell'' cannot make
  money. Minimal number of signals is four: ``buy'', ``sell position'', ``sell short'', ``cover short''.
  The system where e.g. ``buy'' and ``cover short'' is the same signal
  will eventually catastrophically lose money on an event when market go against  position held.
\item Any system entering the position (does not matter long or short)
  during liquidity excess (e.g. $I>I_{IH}$) cannot make money.
  During liquidity excess price movement is
  typically large and ``reverse to the moving average'' type of system
  often use such event as position entering signal.
  The market  after liquidity excess event bounce a little, then typically go to the same direction.
  This give a risk of on what to bet: ``little bounce'' or ``follow the market''.
  What one should do during liquidity excess event is to CLOSE existing position.
  This is very fundamental - if you have a position during market uncertainty -
  eventually you will lose money, you must have ZERO position during
  liquidity excess. This is very important element of the P\&L trading strategy.
\item Any system not entering the position during
  liquidity deficit event (e.g. $I<I_{IL}$) typically lose money.
  Liquidity deficit periods are characterized by small price
  movements and difficult to identify by price-based trading systems.
  Liquidity deficit actually mean that at current price
  buyers and sellers do not match well, and substantial price movement is
  expected. This is very well known by most traders: before large
  market movement volatility (and e.g. standard deviation as its crude measure)
  become very low.
  The direction (whether one should go long or short) during liquidity deficit event
  can, to some extend, be determined by the theory from Section \ref{dynamics}
  and balance of supply--demand generalization (\ref{Ip}).
\item An important issue is to discuss what would happen
  to the markets when this strategy (enter on liquidity deficit, exit
  on liquidity excess) is applied on mass scale by market
  participants. In contrast with other trading strategies,
  which reduce liquidity at current price when applied (when price
  is moved to the uncharted territory the liquidity drains out
  because supply or demand drains out as on classical Fig. \ref{fig:qSD}),
  this strategy actually increase market liquidity at current price.
  This insensitivity to price value is expected to lead
  not to the strategy stopping to work when applied on mass scale
  by market participants, but starting to work better and better
  and to markets destabilization in the end.
\item While proposed theory was developed and tested mostly on
  US equity market, it can be extended to other global markets
  (Treasury, FX, Sovereign Debt, etc)
  with corresponding time scale adjustment. Noticed in Section \ref{volatility}
  similarity between $dv$ and $|dp|$ behavior can probably allow
  the theory to be applied even to the markets, where
  trading volume is not available, using $|dp|$ as a substitute.

\end{itemize}

\appendix
\section{\label{basisses}Non-monomials polynomial bases}
A number of numerical algorithms use monomials basis $x^k$.
However, selection of other bases can be greatly beneficial
to numerical stability improvement.
A choice of a basis satisfying recurrent relation.
\begin{eqnarray}
  Q_{k}(x) &=& (\alpha_{k}x-\delta_k)Q_{k-1}(x) - \gamma_{k} Q_{k−2}(x) \label{rec3}
\end{eqnarray}
has some important stability properties\cite{laurie1979computation,gautschi2004orthogonal}.

For our calculations we use the following four bases:

\begin{itemize}
\item
Laguerre: (see com.polytechnik.utils.Laguerre)
\begin{eqnarray}
  kL_k(x)&=&(2k-1-x)L_{k-1}-(k-1)L_{k-2} \\
  L_0&=&1 \\
  L_{-1}&=& 0
\end{eqnarray}

\item
Legendre: (see com.polytechnik.utils.Legendre)
\begin{eqnarray}
  kP_k&=&x(2k-1)P_{k-1}-(k-1)P_{k-2} \\
  P_0&=&1 \\
  P_{-1}&=& 0
\end{eqnarray}

\item
Chebyshev: (see com.polytechnik.utils.Chebyshev)
\begin{eqnarray}
  T_k&=&2x T_{k-1}-T_{k-2} \\
  T_0&=&1 \\
  T_{1}&=& x
\end{eqnarray}

\item
Hermite (actually $He$ basis): (see com.polytechnik.utils.HermiteE)
\begin{eqnarray}
  H_k&=&x H_{k-1}-(k-1)H_{k-2} \\
  H_0&=&1 \\
  H_{-1}&=& 0
\end{eqnarray}
\end{itemize}

To use these bases in calculations we need to
be able to perform standard operations on polynomials in these bases
$F(x)=\sum_{k=0}^{k=n} f_kQ_k(x)$, where $f_k$ is now the coefficient by $Q_k$,
not by $x^k$ as in monomial basis.

\begin{enumerate}
\item Multiplication operation:
\begin{eqnarray}
  Q_i Q_j&=&\sum_{k=0}^{k=i+j}c_k^{ij}Q_k \label{cmul}
\end{eqnarray}
For the four mentioned bases the coefficients $c_k^{ij}$ from (\ref{cmul}) are known:
For Laguerre basis: Ref. \onlinecite{watson}.
For Legendre Basis: Ref. \onlinecite{gradshtein}, formulae 8.915.5, A(9036), page 1040.
For Chebyshev Basis: Ref. \onlinecite{milne1972handbook}, formulae 22.7.24, p. 872.
For Hermite Basis: Ref. \onlinecite{Carlitz} or \onlinecite{Prudnikov} formulae 4.5.1.11 page 569.

\item Multiplication by $ax+b$. Use 3 term recurrence relation.

\item Given a set of observations $x_j$ and $w_j$ calculate the moments as $\sum_j Q_n(x_j)w_j$.
  Use 3 term recurrence relation (see the method calculateMomentsFromSample).

\item Expand $ax+b$ argument $Q_n(ax+b)=\sum_{j=0}^{j=n}d^{(n)}_jQ_j(x)$. Use 3 term recurrence relation
to find $d^{(n)}_j$.

\item Synthetic division. For a given polynomial $P=\sum_{k=0}^{k=n_p} p_kQ_k$ and $D=\sum_{k=0}^{k=n_d} d_k Q_k$
find polynomials $R$ and $Q$ such as $P=Q*D+R$.
For $n_d=1$ result can be calculated directly from three term recurrence (\ref{rec3}),
for $n_d>1$  use
the (\ref{cmul}) coefficients and solve linear system with respect to $R$ and $Q$ coefficients.

\item Calculation of $\sum_{k=0}^{k=n} f_kQ_k(x)$ at $x$.
Use Clenshaw recurrence formula see Ref. \onlinecite{fox1968chebyshev}, page 56.

\item Integration and differentiation of a function $\sum_{k=0}^{k=n} f_kQ_k(x)$ at $x$.
Use $Q_k(x)$ integration and differentiation formulas from  Refs. \onlinecite{gradshtein,milne1972handbook,Prudnikov},
then apply Clenshaw recurrence formula.

\item Given $F(x)=\sum_{k=0}^{k=n} f_kQ_k(x)$ find the roots(possibly complex)
of $F(x)=0$. Build confederate matrix\cite{Barnett,Bella}
the eigenvalues of which give the polynomial $F(x)$ roots.
See  getConfederateMatrix(final double [] coefs) method and
 com. polytechnik. utils. PolynomialRootsConfederateMatrixABasis class.

\end{enumerate}

We have a numerical library implementing these (and also some other)
polynomial operations for the four bases in question
(see mentioned above four classes extending the com. polytechnik. utils. BasisPolynomials).
The code is availble from authors\cite{polynomialcode}.
To show simple application of these bases let us apply them to quadratures
calculations. This will be demonstrated in Appendix \ref{quadratures}

\section{\label{quadratures}Quadratures calculation}
In this section given the moments $<Q_k>_{\mu}$ we
apply the operations from Appendix \ref{basisses}
to calculate Gauss, Radau, Kronrod and Multiple Orthogonality quadratures.

Gaussian quadratures.
Using multiplication coefficients (\ref{cmul})
obtain matrices $M_{\mu}[x]$ and $M_{\mu}[1]$.
The first one is obtained initially by multiplication by $x$,
then using (\ref{cmul}), the second one is obtained by direct
application of (\ref{cmul}). Solve generalized eigenvalues problem
\begin{eqnarray}
  M_{\mu}[x]|\psi>&=&\lambda M_{\mu}[1]|\psi>
  \label{efXquadratures}
\end{eqnarray}
The eigenvalues are the quadrature nodes $x_k$, $k=[0..n]$
and the weights $w_k=1/\left(\psi^{(k)}(x_k)\right)^2$, where
$\psi^{(k)}(x_k)$ is the value of $k$-th eigenfunction at $x_k$,
which is $\sum_{j=0}^{j=n} \psi^{(k)}_jQ_j(x_k)$. Because
$<\psi^{(j)}|M_{\mu}[1]|\psi^{(k)}>=\delta_{jk}$ 
the $K(x,y,\mu)$ from (\ref{vKrepr}) and corresponding
Christoffel function has a very simple form
in $|\psi^{(k)}>$ basis:
\begin{eqnarray}
  K(x,y,\mu)&=&\sum_{k=0}^{k=n} \psi^{(k)}(x)\psi^{(k)}(y)
  \label{Kpsibasis}
\end{eqnarray}
The $\psi^{(k)}(x)$ are equal(within a constant) to the
Lagrange interpolating polynomial
 built on  $x_j$, $j=[0..n]$ nodes ($\psi^{(k)}(x_j)=0$ for $j\ne k $).

Radau quadratures.
Using multiplication coefficients (\ref{cmul})
obtain matrices $M_{\mu}[(x_0-x)x]$ and $M_{\mu}[(x_0-x)]$,
then use Gaussian quadratures for nodes and weights calculation.

Kronrod quadratures\cite{kronrod1964,laurie1997calculation}.
Build Gaussian quadrature first, then obtain $n$-th orthogonal
polynomial $P_n$ on measure $\mu$ (e.g. by multiplication  $\psi^{(k)}$ by $x-x_k$
or in some other way), then calculate $<Q_k P_n>$ moments using (\ref{cmul})
and calculate Gaussian quadrature once again on these new moments.
If successful - the result give Kronrod nodes. Once Kronrod nodes are known
Kronrod weights can be easily calculated from first and second Gaussian quadrature weights.
See the code in com.polytechnik.utils. OrthogonalPolynomialsABasis. getKronrodQuadratures.

Multiple orthogonality\cite{borges1994class}. See the code in
com.polytechnik.utils. OrthogonalPolynomialsABasis. getQuadraturesForMultipleOrthogonalPolynomial

Java code
is available from authors \cite{polynomialcode}.

\section{\label{ggaussqusage}Distribution Parameters Estimation with Gaussian Quadratures}
The quadratures we have built in Appendix \ref{quadratures}
can be applied for distribution parameters estimation.
For a positively defined measure $d\mu$ with existing moments $[0..2n-1]$
it is possible to build $n$-point quadrature rule,
such that the relation
\begin{eqnarray}
  \left< \Pi(x) \right> &=& \sum_{k=1}^{k=n}\Pi(x_k)\omega_k
  \label{quadraddef}
\end{eqnarray}
is exact if $\Pi(x)$ is arbitrary polynomial of degree $2n-1$ or less.
The nodes $x_k$ and the weights $\omega_k$
define Gaussian quadrature\cite{gautschi2004orthogonal,totik,szego1974orthogonal}
While most quadrature applications focus on using (\ref{quadraddef})
for integrals estimation, 
it can be viewed as interpolation of the measure $d\mu$ itself
by a discrete measure with support on quadrature nodes,
i.e. by delta functions at points $x_k$ and magnitude $\omega_k$.
(See the Ref. \cite{totik} for distribution of $x_k$ (the roots of the $n$-th order orthogonal
polynomial with respect to measure $d\mu$) review in various cases).
Some trivial usage of a quadrature can be an estimation of a quantiles
(e.g. median) of the measure $d\mu$ using
the discrete measure $\omega_k$  as a substitute.

In this appendix we present a new skewness  estimator for a distribution.
Given the $\left<Q_k\right>; k=0,1,2,3$ moments it is possible to build two point 
quadrature rule. Assuming the quadrature nodes are ordered in ascending order $x_1<x_2$
define the skewness as asymmetry of nodes weights 
\begin{eqnarray}
\Gamma&=&\frac{\omega_1-\omega_2}{\omega_1+\omega_2} \label{skewness}
\end{eqnarray}
The definition (\ref{skewness}) is bounded to $[-1;1]$ interval
because all $\omega_k$ are positive.

\begin{figure}
\includegraphics[width=10cm]{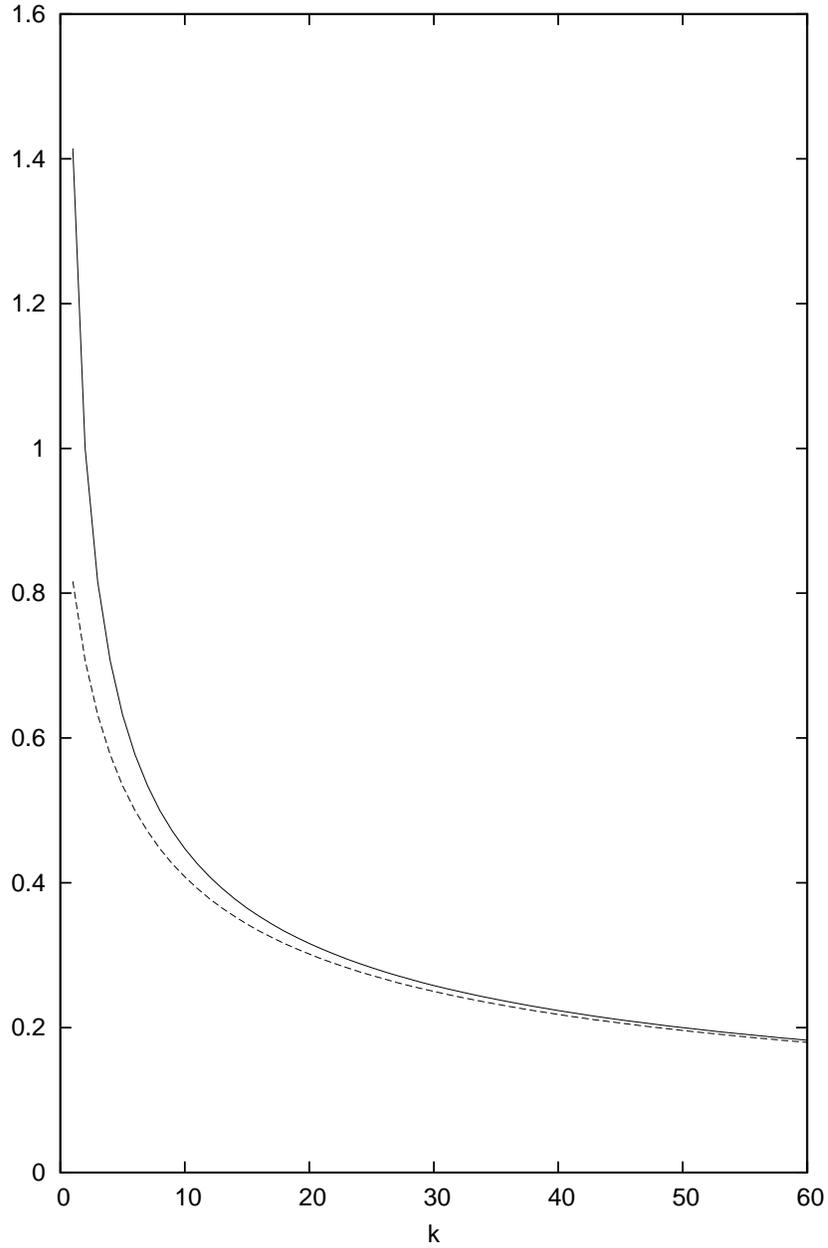}
\caption{\label{chi2} Skewness for chi--squared distribution.
Solid line: half of regular skewness $\sqrt{8/k}$. Dashed line: modified skewness from
(\ref{skewness2}) }
\end{figure}

Practical Gaussian quadrature calculation can be rather complicated for a large $n$ 
because of numerical instability,
but for $n=2$ calculation is trivial and can be performed even in monomials basis.
Consider $L^2$ extremal problem of $\int (a+bx+x^2)^2 d\mu$, what lead to linear system 
and the values for $a$ and $b$ are:
\begin{eqnarray}
d&=&\big<x^2\big>\big<1\big>-\big<x\big>^2\\
a&=&\left(\big<x^3\big>\big<x\big>-\big<x^2\big>^2\right)/d \\
b&=&\left(\big<x^2\big>\big<x\big>-\big<x^3\big>\big<1\big>\right)/d
\end{eqnarray}
then the nodes are the $a+bx+x^2$ roots and the weights are:
\begin{eqnarray}
x_{1,2}&=&\frac{-b\pm\sqrt{b^2-4a}}{2} \\
\omega_1&=&\left<1\right>\frac{\overline{x}-x_2}{x_1-x_2} \\
\omega_2&=&\left<1\right>\frac{x_1-\overline{x}}{x_1-x_2}
\end{eqnarray}
Then (\ref{skewness}) becomes
\begin{eqnarray}
\Gamma&=&\frac{2\overline{x}-x_1-x_2}{x_1-x_2}
=-\frac{2\overline{x}+b}{\sqrt{b^2-4a}}
 \label{skewness2} \\
\Gamma_{[x]}&=&(x_1+x_2)/2-\overline{x}
\label{skewness2x}
\end{eqnarray}
The skewness defined in (\ref{skewness}) and calculated in (\ref{skewness2}) is very similar 
to regular skewness $\gamma_1$ from (\ref{gamma1}) when applied to commonly used distributions.
\begin{eqnarray}
\overline{x}&=&\left<x\right>/\left<1\right> \\
\sigma^2&=&\frac{\left<(x-\overline{x})^{2}\right>}{\left<1\right>} \\
\gamma_1&=&\frac{\left<(x-\overline{x})^{3}\right>}{\left<1\right>\sigma^3} \label{gamma1}
\end{eqnarray}
On Fig. \ref{chi2} a plot of regular skewness $\gamma_1$ from (\ref{gamma1}) and
``modified'' skewness $\Gamma$ from (\ref{skewness}) are presented for chi--squared distribution 
as a function of degree of freedom $k$ (regular skewness is equal to exactly $\sqrt{8/k}$, on a chart it is divided by two to have the same asymptotic as (\ref{skewness2}) at $k\to\infty$).
In some situations a definition of skewness, having the 
dimension of $x$ is required (e.g. a difference between mean and median
used in nonparametric skew).
For such estimation half of (\ref{skewness2}) nominator can be used,
what gives (\ref{skewness2x}) as a difference between 
the midpoint of $x_1$ and $x_2$
and mean $\overline{x}$. Note that the $\overline{x}$ is the root
of first order orthogonal polynomial $P_1(x)$ built on $d\mu$ 
and $x_1$ and $x_2$ are the roots of the second
order orthogonal polynomial $P_2(x)$ built on $d\mu$,
thus the (\ref{skewness2x}) is a
difference between a midpoint of $P_2(x)$ roots and $P_1(x)$ root.
See the com.polytechnik.utils.Skewness for the code calculating $\Gamma$
from the $[0..3]$ moments in arbitrary basis.

First $[0..2n-1]$ moments of a positive measure
can be one--to-one mapped to $n$-point Gaussian quadrature.
A modified skewness estimation as asymmetry
of two--point Gaussian quadrature weighs is proposed. 
This modified skewness has additional important properties,
such as bounded to $[-1..1]$ interval 
and being applicable well to two-mode distribution,
it gives exact answer, for example, in case 
of discrete distribution with two support points.
Note, that discrete distribution is a typical problematic case for skewness estimation\cite{von2005mean}.
While quadratures approach can be easily applied to skewness estimation,
kurtosis estimation from Gaussian quadrature is not possible if input moments
are limited to the same ones used in classic definition of kurtosis.
Classic kurtosis estimation requires $[0..4]$ moments
for estimation, but 3-point Gaussian quadrature requires $[0..5]$ moments.
In this sense quadrature--based skewness estimation is some kind special,
because it can be built using the same input moments as classically defined skewness. In practical applications the Christoffel function (\ref{vKrepr}) asymptotic $1/K(x,x,\mu)$
can be much more successfully, than kurtosis, applied for testing a distribution on ``fat tails''. Technically  Christoffel function behavior
can be better understood
in the (\ref{efXquadratures}) eigenfunctions basis
(in which $K(x,y,\mu)$ has a very simple form (\ref{Kpsibasis}))
rather than in the original $Q_k(x)$ basis, in which $K(x,y,\mu)$
has a general form (\ref{Krepr}). Given distribution sample
to obtain $K(x,x,\mu)$ select a basis out of four bases considered
(for numerical stability choose the one the measure of which is most similar
to distribution of the sample and scale $x$ to the basis measure support),
then use basis implementation
of com.polytechnik.utils. BasisPolynomials. calculateMomentsFromSample
to obtain $<Q_k>$ moments, after that make the $M[1]$ matrix using
com.polytechnik.utils. OrthogonalPolynomialsABasis .getQQMatr,
inverse it (obtain $G^{-1}$) by applying e.g. com.polytechnik.utils .Linsystems .getInvertedMatrix,
and finally calculate the polynomail $\mathbf{Q}(x)G^{-1}\mathbf{Q}(x)$
by using the
com.polytechnik.utils. OrthogonalPolynomialsABasis .getKK.
Java code for e.g. Chebyshev basis would look about like this:
\begin{verbatim}
/** The method calculates K(x,x) (2d-1 elements returned, the polynomial of 2d-2 order) 
  * in Chebyshev basis from observations sample x[].
  */
static double [] getKxxFromSample(final int d,final double [] x){
final com.polytechnik.utils.OrthogonalPolynomialsABasis Q=
       new com.polytechnik.utils.OrthogonalPolynomialsChebyshevBasis();
return Q.getKK(d,com.polytechnik.utils.Linsystems.getInvertedMatrix(d,d,
           Q.getQQMatr(d,Q.B.calculateMomentsFromSample(2*d-1-1,x))),d);
}
\end{verbatim}

\section{\label{runge}Runge Oscillations supression}
We take Runge function
\begin{eqnarray}
  f(x)&=&\frac{1}{1+25x^2} \label{rungeF}
\end{eqnarray}
And interpolate it on $[-1;1]$ interval choosing the measure $d\mu=dx$
and $n=6$. 
$A_{LS}(x)$ is least square approximation (\ref{pinterp})
and $A_{RN}(x)$ Radon--Nikodym approximation (\ref{RN}).
\begin{eqnarray}
  G_{ij}&=&\int\limits_{-1}^{1}dx Q_i(x)Q_j(x) = <Q_i(x)Q_j(x)> =M_{ij}[1] \\ 
  A_{LS}(x)&=&Q_i(x) G^{-1}_{ij}<Q_j(x)f> = \mathbf{Q}(x)G^{-1} <\mathbf{Q}f>
  \label{Apprleastsq} \\
  A_{RN}(x)&=&\frac{Q_i(x) G^{-1}_{ij} <Q_jQ_kf> G^{-1}_{kl} Q_l(x)}
  {Q_i(x) G^{-1}_{ij}Q_j(x)}=
  \frac{\mathbf{Q}(x)G^{-1}M[f]G^{-1}\mathbf{Q}(x)}{\mathbf{Q}(x)G^{-1}M[1]G^{-1}\mathbf{Q}(x)}
  \label{RNsimple}
\end{eqnarray}

\begin{figure}
\includegraphics[width=10cm]{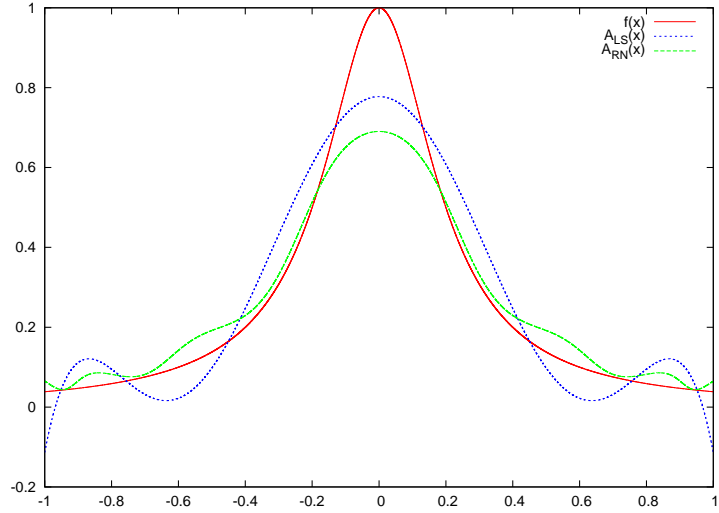}
\caption{\label{fig:fig_runge}
  Original Runge function $f$, Least Squares approximation $A_{LS}$
  and Radon--Nikodym approximation $A_{RN}$
}  
\end{figure}

The results are presented on Fig. \ref{fig:fig_runge}.
One can see that near edges oscillations are much
less severe, when  Radon--Nikodym approximation as polynomials ratio is used
for the interpolation of $f$. One can see from the chart typical
behavior difference for least square and Radon--Nikodym approximations:
Least squares have diverging oscillations near
measure support boundaries and tend to infinity with the distance to
measure support increase. Radon--Nikodym have converging
oscillations near
measure support boundaries and tend to a constant with the distance to
measure support increase.
The code calculating $A_{RN}(x)$ from
$<Q_k>$ and $<fQ_k>$ moments
is very similar to an example
calculating the $K(x,x,\mu)$ polynomial 
at the end of Appendix \ref{ggaussqusage}, with the difference
that the calculations now have to be performed twice: first time for denominator,
what give exactly $K(x,x,\mu)$, and second time for nominator,
with only difference is that instead of the matrix $G^{-1}$
the matrix $G^{-1}M[f]G^{-1}$ should be used.
See the com.polytechnik.utils. NevaiOperator. getNevaiOperator
as an example where these calculations are implemented
and the polynomials for nominator and denominator are calculated
from the moments in a given basis.

Another, worth to mention point, is related to derivatives calculation.
For this the moments $\left<Q_k df/dx\right>$
should be calculated first (for the measures like (\ref{muflaguerre}) or (\ref{muslegendre}) this can be done using $\left<Q_k f\right>$ moments and integration by parts),
and only then applying Radon--Nikodym approximation like (\ref{RNsimple}) using the derivative moments.
If one, instead of using the $\left<Q_k df/dx\right>$ moments,
would  differentiate $f$ approximation expression (\ref{RNsimple}) directly --
the result will be incorrect.

\section{\label{spur}Matrix averages}

In the beginning of Section \ref{kin} we mentioned an effective way
of average and correlation calculation. Specifically we need an effective
way to calculate $<fg>$ given only $<Q_k f>$ and $<Q_k g>$ information.
The approach mentioned in Section \ref{kin} is actually
\begin{eqnarray}
  G_{ij}&=&<Q_i Q_j>  \\
  \overline{fg}&=&\frac{<\mathbf{Q} f> G^{-1}<\mathbf{Q} g>}
  {<\mathbf{Q}> G^{-1}<\mathbf{Q}>}
  \label{fgaver}
\end{eqnarray}
(in this appendix
we mix vector $<\mathbf{Q}f>$ and index $<Q_k f>$ notations for notation
compactness, but this should not mislead the reader).

The expression(\ref{fgaver}) can be also considered as conversion
of $f(t)$ and $g(t)$ timeseries to vectors $<\mathbf{Q} f>$ and $<\mathbf{Q} g>$
then taking inner product of them with matrix $G^{-1}$  defining inner product (another way to look at this is to consider least squares
approximation of $f(x)$ and $g(x)$ then taking average of two interpolated functions product).

In a way how Radon--Nikodym derivatives improve interpolation
of a function, the transition from a vector $<Q_k f>$ 
to matrix $M[f]$ can similary improve calculations of an average.
Let us use the $M_{ij}[f]=<Q_i f Q_j>$ from (\ref{mc}) and define an average $\overline{f}$:
\begin{eqnarray}
  \overline{f}&=&\frac{\mathrm{Spur}\left(G^{-1}M[f]\right)}{\dim G} \label{spuraver}
\end{eqnarray}
where $\mathrm{Spur}$ is matrix trace (sum of diagonal elements) operator.
It is easy to see that the definition (\ref{spuraver})
immediately give (note that $G=M[1]$ and $\mathrm{Spur}\left(G^{-1}M[f]\right)=\mathrm{Spur}\left(M[f]G^{-1}\right)$)
\begin{eqnarray}
  \dim G&=&\mathrm{Spur}\left(G^{-1}G\right)=n+1 \\
  \overline{fg}&=&\frac{\mathrm{Spur}\left(G^{-1}M[f]G^{-1}M[g]\right)}{\dim G} \label{fgspuraver}
\end{eqnarray}
The average (\ref{fgspuraver}) is related
to quantum mechanics \cite{malyshkin2015norm}
density matrix --type of average, 
and it  has all the regular
properties of average, but operates on matrices
(an equivalent of quantum mechanics density matrix), not on vectors.
This greatly increase stability of calculations
(both because of using more moments $[0..2n]$ instead of $[0..n]$
and because of matrix nature of the expression (\ref{spuraver})).
If the basis $Q_k(x)$ is chosen in a way the $G$ is a unit matrix
 then all $G^{-1}$ terms vanish and
$\overline{f}$ is just $\mathrm{Spur}(M[f])/\dim G$ and $\overline{fg}$ is just $\mathrm{Spur}(M[f]M[g])/\dim G$.
Interesting properties arise when matrices $M[f]$ and $M[g]$
have some special properties (e.g. have common basis in which both are diaginal,
commutate, etc.).

Note, that the formulae (\ref{fgspuraver}) practically allows
to calculate stock cross correlation in linear time.
To obtain price covariance of any two stocks $p$ and $q$:
obtain $M[p]$ and $M[q]$ matrices (\ref{mc}) from $[0..2n]$ moments of $p$ and $q$ timeseries,
then use the (\ref{fgspuraver}) for $\overline{pq}-\overline{p}\,\,\overline{q}$.

\bibliography{LD}

\begin{thebibliography}{39}%
\makeatletter
\providecommand \@ifxundefined [1]{%
 \@ifx{#1\undefined}
}%
\providecommand \@ifnum [1]{%
 \ifnum #1\expandafter \@firstoftwo
 \else \expandafter \@secondoftwo
 \fi
}%
\providecommand \@ifx [1]{%
 \ifx #1\expandafter \@firstoftwo
 \else \expandafter \@secondoftwo
 \fi
}%
\providecommand \natexlab [1]{#1}%
\providecommand \enquote  [1]{``#1''}%
\providecommand \bibnamefont  [1]{#1}%
\providecommand \bibfnamefont [1]{#1}%
\providecommand \citenamefont [1]{#1}%
\providecommand \href@noop [0]{\@secondoftwo}%
\providecommand \href [0]{\begingroup \@sanitize@url \@href}%
\providecommand \@href[1]{\@@startlink{#1}\@@href}%
\providecommand \@@href[1]{\endgroup#1\@@endlink}%
\providecommand \@sanitize@url [0]{\catcode `\\12\catcode `\$12\catcode
  `\&12\catcode `\#12\catcode `\^12\catcode `\_12\catcode `\%12\relax}%
\providecommand \@@startlink[1]{}%
\providecommand \@@endlink[0]{}%
\providecommand \url  [0]{\begingroup\@sanitize@url \@url }%
\providecommand \@url [1]{\endgroup\@href {#1}{\urlprefix }}%
\providecommand \urlprefix  [0]{URL }%
\providecommand \Eprint [0]{\href }%
\providecommand \doibase [0]{http://dx.doi.org/}%
\providecommand \selectlanguage [0]{\@gobble}%
\providecommand \bibinfo  [0]{\@secondoftwo}%
\providecommand \bibfield  [0]{\@secondoftwo}%
\providecommand \translation [1]{[#1]}%
\providecommand \BibitemOpen [0]{}%
\providecommand \bibitemStop [0]{}%
\providecommand \bibitemNoStop [0]{.\EOS\space}%
\providecommand \EOS [0]{\spacefactor3000\relax}%
\providecommand \BibitemShut  [1]{\csname bibitem#1\endcsname}%
\let\auto@bib@innerbib\@empty
\bibitem [{\citenamefont {Mandelbrot}\ and\ \citenamefont
  {Hudson}(2014)}]{mandelbrot2014misbehavior}%
  \BibitemOpen
  \bibfield  {author} {\bibinfo {author} {\bibfnamefont {Benoit}\ \bibnamefont
  {Mandelbrot}}\ and\ \bibinfo {author} {\bibfnamefont {Richard~L}\
  \bibnamefont {Hudson}},\ }\href@noop {} {\emph {\bibinfo {title} {The
  Misbehavior of Markets: A fractal view of financial turbulence}}}\ (\bibinfo
  {publisher} {Basic books},\ \bibinfo {year} {2014})\BibitemShut {NoStop}%
\bibitem [{\citenamefont {Corcoran}(2007)}]{corcoran2007long}%
  \BibitemOpen
  \bibfield  {author} {\bibinfo {author} {\bibfnamefont {Clive~M}\ \bibnamefont
  {Corcoran}},\ }\href@noop {} {\emph {\bibinfo {title} {Long/short market
  dynamics: trading strategies for today's markets}}},\ Vol.\ \bibinfo {volume}
  {323}\ (\bibinfo  {publisher} {John Wiley \& Sons},\ \bibinfo {year}
  {2007})\BibitemShut {NoStop}%
\bibitem [{\citenamefont {McCauley}(2009)}]{mccauley2009dynamics}%
  \BibitemOpen
  \bibfield  {author} {\bibinfo {author} {\bibfnamefont {Joseph~L}\
  \bibnamefont {McCauley}},\ }\href@noop {} {\emph {\bibinfo {title} {Dynamics
  of markets: The new financial economics}}}\ (\bibinfo  {publisher} {Cambridge
  University Press},\ \bibinfo {year} {2009})\BibitemShut {NoStop}%
\bibitem [{\citenamefont {Choi}\ \emph {et~al.}(2015)\citenamefont {Choi},
  \citenamefont {Larsen},\ and\ \citenamefont {Seppi}}]{choi2015information}%
  \BibitemOpen
  \bibfield  {author} {\bibinfo {author} {\bibfnamefont {Jin~Hyuk}\
  \bibnamefont {Choi}}, \bibinfo {author} {\bibfnamefont {Kasper}\ \bibnamefont
  {Larsen}}, \ and\ \bibinfo {author} {\bibfnamefont {Duane~J}\ \bibnamefont
  {Seppi}},\ }\bibfield  {title} {\enquote {\bibinfo {title} {Information and
  trading targets in a dynamic market equilibrium},}\ }\href
  {http://arxiv.org/abs/1502.02083} {\bibfield  {journal} {\bibinfo  {journal}
  {arXiv preprint arXiv:1502.02083}\ } (\bibinfo {year} {2015})}\BibitemShut
  {NoStop}%
\bibitem [{\citenamefont {Malyshkin}\ and\ \citenamefont
  {Bakhramov}(2015)}]{2015arXiv151005510G}%
  \BibitemOpen
  \bibfield  {author} {\bibinfo {author} {\bibfnamefont
  {Vladislav~Gennadievich}\ \bibnamefont {Malyshkin}}\ and\ \bibinfo {author}
  {\bibfnamefont {Ray}\ \bibnamefont {Bakhramov}},\ }\bibfield  {title}
  {\enquote {\bibinfo {title} {{Mathematical Foundations of Realtime Equity
  Trading. Liquidity Deficit and Market Dynamics. Automated Trading
  Machines.}}}\ }\href {http://arxiv.org/abs/1510.05510} {\bibfield  {journal}
  {\bibinfo  {journal} {ArXiv e-prints}\ } (\bibinfo {year} {2015})},\ \bibinfo
  {note} {\url{http://arxiv.org/abs/1510.05510}},\ \Eprint
  {http://arxiv.org/abs/1510.05510} {arXiv:1510.05510 [q-fin.CP]} \BibitemShut
  {NoStop}%
\bibitem [{\citenamefont {Hautsch}\ and\ \citenamefont
  {Huang}(2011)}]{nasdaqord}%
  \BibitemOpen
  \bibfield  {author} {\bibinfo {author} {\bibfnamefont {Nikolaus}\
  \bibnamefont {Hautsch}}\ and\ \bibinfo {author} {\bibfnamefont {Ruihong}\
  \bibnamefont {Huang}},\ }\href
  {http://sfb649.wiwi.hu-berlin.de/papers/pdf/SFB649DP2011-056.pdf} {\enquote
  {\bibinfo {title} {Limit order flow, market impact and optimal order sizes:
  Evidence from nasdaq totalview-itch data},}\ } (\bibinfo {year}
  {2011})\BibitemShut {NoStop}%
\bibitem [{\citenamefont {OMX}(2014)}]{itchfeed}%
  \BibitemOpen
  \bibfield  {author} {\bibinfo {author} {\bibfnamefont {Nasdaq}\ \bibnamefont
  {OMX}},\ }\href
  {http://www.nasdaqtrader.com/content/technicalsupport/specifications/dataproducts/nqtv-itch-v4_1.pdf}
  {\emph {\bibinfo {title} {NASDAQ TotalView-ITCH 4.1}}},\ \bibinfo {type}
  {Report}\ (\bibinfo  {institution} {Nasdaq OMX},\ \bibinfo {year} {2014})\
  \bibinfo {note} {{Also see data files samples
  \url{ftp://emi.nasdaq.com/ITCH/}}}\BibitemShut {NoStop}%
\bibitem [{\citenamefont {Lynch}(29~Apr.~2014)}]{secchair}%
  \BibitemOpen
  \bibfield  {author} {\bibinfo {author} {\bibfnamefont {Sarah~N.}\
  \bibnamefont {Lynch}},\ }\href
  {http://www.reuters.com/article/2014/04/29/us-sec-highspeed-trading-idUSBREA3S0OO20140429}
  {\enquote {\bibinfo {title} {Sec chair to congress: 'the markets are not
  rigged'},}\ } (\bibinfo {year} {29~Apr.~2014})\BibitemShut {NoStop}%
\bibitem [{Not()}]{NoteBookInfo}%
  \BibitemOpen
  \href@noop {} {}\bibinfo {note} {There is a singnificant effort we made
  trying to use book informationation to predict market dynamics. The
  information we tried to use without much success: volume disbalance near book
  edges, spread, cancellation rate, cancellation from best price, and many
  other. The only more or less useful order book information was origination
  time of the order to be executed. When orders on best level are old enough
  (old orders are near book edge) this typically indicates liquidity deficit
  event and is a good indication of stopping trading near this price. (This is
  also true for an executed trade, even without book data available: if
  recently executed orders were put to order book long time ago this means that
  at current price level there are no new liqudity available, because no recent
  orders get matched at current price.) A usage of order book information
  related to limit orders volume, was not successful at all. For example use
  order book volume $dv$ as measure weight use Christoffel function
  $w(p)=\frac{1}{K(p,p)}=\frac{1}{\pi_k(p)\left(<\pi(p)\pi(p)>\right)^{-1}_{kl}\pi_l(p)}$
  for both buy and sell sides of the order book. This is effectively an
  interpolation of book volume. The result show that for liquid stock one side
  is significantly large than other, the volume fluctuates, no good results are
  received from matching interpolated $w(p)$ for buy and sell sides:
  $w_{sell}(p)=w_{buy}(p)$ in an attempt to find an equilibrium $p$. The price
  differ significantly from buy/sell book edges. Using Haar measure instead of
  $dv$ gave no improvement also.}\BibitemShut {Stop}%
\bibitem [{\citenamefont {Moro}\ \emph {et~al.}(03~Aug.~2009)\citenamefont
  {Moro}, \citenamefont {Vicente}, \citenamefont {Moyano}, \citenamefont
  {Gerig}, \citenamefont {Farmer}, \citenamefont {Vaglica}, \citenamefont
  {Lillo},\ and\ \citenamefont {Mantegna}}]{farmerimpact}%
  \BibitemOpen
  \bibfield  {author} {\bibinfo {author} {\bibfnamefont {Esteban}\ \bibnamefont
  {Moro}}, \bibinfo {author} {\bibfnamefont {Javier}\ \bibnamefont {Vicente}},
  \bibinfo {author} {\bibfnamefont {Luis~G.}\ \bibnamefont {Moyano}}, \bibinfo
  {author} {\bibfnamefont {Austin}\ \bibnamefont {Gerig}}, \bibinfo {author}
  {\bibfnamefont {J.~Doyne}\ \bibnamefont {Farmer}}, \bibinfo {author}
  {\bibfnamefont {Gabriella}\ \bibnamefont {Vaglica}}, \bibinfo {author}
  {\bibfnamefont {Fabrizio}\ \bibnamefont {Lillo}}, \ and\ \bibinfo {author}
  {\bibfnamefont {Rosario~N.}\ \bibnamefont {Mantegna}},\ }\bibfield  {title}
  {\enquote {\bibinfo {title} {Market impact and trading profile of large
  trading orders in stock markets},}\ }\href {http://arxiv.org/abs/0908.0202}
  {\bibfield  {journal} {\bibinfo  {journal} {arXiv:0908.0202 [q-fin.TR]}\ }
  (\bibinfo {year} {03~Aug.~2009})}\BibitemShut {NoStop}%
\bibitem [{\citenamefont {Yakushev}(2009)}]{EyakushevComm}%
  \BibitemOpen
  \bibfield  {author} {\bibinfo {author} {\bibfnamefont {Eugeny}\ \bibnamefont
  {Yakushev}},\ }\href@noop {} {} (\bibinfo {year} {2009}),\ \bibinfo {note}
  {private communication.}\BibitemShut {Stop}%
\bibitem [{\citenamefont {Gautschi}(2004)}]{gautschi2004orthogonal}%
  \BibitemOpen
  \bibfield  {author} {\bibinfo {author} {\bibfnamefont {Walter}\ \bibnamefont
  {Gautschi}},\ }\href@noop {} {\emph {\bibinfo {title} {Orthogonal
  polynomials: computation and approximation}}}\ (\bibinfo  {publisher} {Oxford
  University Press on Demand},\ \bibinfo {year} {2004})\BibitemShut {NoStop}%
\bibitem [{\citenamefont {Malyshkin}(2016)}]{2016arXiv160204423G}%
  \BibitemOpen
  \bibfield  {author} {\bibinfo {author} {\bibfnamefont
  {Vladislav~Gennadievich}\ \bibnamefont {Malyshkin}},\ }\bibfield  {title}
  {\enquote {\bibinfo {title} {{Market Dynamics. On Supply and Demand
  Concepts}},}\ }\href {http://arxiv.org/abs/1602.04423} {\bibfield  {journal}
  {\bibinfo  {journal} {ArXiv e-prints}\ } (\bibinfo {year} {2016})},\ \bibinfo
  {note} {\url{http://arxiv.org/abs/1602.04423}},\ \Eprint
  {http://arxiv.org/abs/1602.04423} {arXiv:1602.04423} \BibitemShut {NoStop}%
\bibitem [{\citenamefont {Malyshkin}(2009)}]{malha}%
  \BibitemOpen
  \bibfield  {author} {\bibinfo {author} {\bibfnamefont {Gennadii~Stepanovich}\
  \bibnamefont {Malyshkin}},\ }\href
  {http://www.elektropribor.spb.ru/publ/rbook32} {\emph {\bibinfo {title}
  {Optimal and Adaptive Methods of Hydroacoustic Signal Processing. Vol 1.
  Optimal methods. (in Russian).}}}\ (\bibinfo  {publisher} {Elektropribor
  Publishing},\ \bibinfo {year} {2009})\ \bibinfo {note} {{ISBN:}
  978-5-900780-90-0}\BibitemShut {NoStop}%
\bibitem [{\citenamefont {Totik}(11~Nov.~2005)}]{totik}%
  \BibitemOpen
  \bibfield  {author} {\bibinfo {author} {\bibfnamefont {Vilmos}\ \bibnamefont
  {Totik}},\ }\bibfield  {title} {\enquote {\bibinfo {title} {Orthogonal
  polynomials},}\ }\href {http://www.math.technion.ac.il/sat/papers/3/3.pdf}
  {\bibfield  {journal} {\bibinfo  {journal} {Surveys in Approximation Theory}\
  }\textbf {\bibinfo {volume} {1}},\ \bibinfo {pages} {70--125} (\bibinfo
  {year} {11~Nov.~2005})}\BibitemShut {NoStop}%
\bibitem [{\citenamefont {Kolmogorov}\ and\ \citenamefont
  {Fomin}(8~May~2012)}]{kolmogorovFA}%
  \BibitemOpen
  \bibfield  {author} {\bibinfo {author} {\bibfnamefont {A.~N.}\ \bibnamefont
  {Kolmogorov}}\ and\ \bibinfo {author} {\bibfnamefont {S.~V.}\ \bibnamefont
  {Fomin}},\ }\href@noop {} {\emph {\bibinfo {title} {Elements of the Theory of
  Functions and Functional Analysis}}}\ (\bibinfo  {publisher} {Martino Fine
  Books (May 8, 2012)},\ \bibinfo {year} {8~May~2012})\BibitemShut {NoStop}%
\bibitem [{\citenamefont {Taleb}(2014)}]{taleb2014silent}%
  \BibitemOpen
  \bibfield  {author} {\bibinfo {author} {\bibfnamefont {Nassim~Nicholas}\
  \bibnamefont {Taleb}},\ }\bibfield  {title} {\enquote {\bibinfo {title}
  {Silent risk: Lectures on fat tails,(anti) fragility, and asymmetric
  exposures},}\ }\href {http://www.fooledbyrandomness.com/FatTails.html}
  {\bibfield  {journal} {\bibinfo  {journal} {Available at SSRN}\ } (\bibinfo
  {year} {2014})}\BibitemShut {NoStop}%
\bibitem [{\citenamefont {Simon}(2011)}]{BarrySimon}%
  \BibitemOpen
  \bibfield  {author} {\bibinfo {author} {\bibfnamefont {Barry}\ \bibnamefont
  {Simon}},\ }\href@noop {} {\emph {\bibinfo {title} {Szeg\H{o}’s Theorem and
  Its Descendants}}}\ (\bibinfo  {publisher} {Princeton University Press},\
  \bibinfo {year} {2011})\BibitemShut {NoStop}%
\bibitem [{\citenamefont {Nevai}(1986)}]{nevai}%
  \BibitemOpen
  \bibfield  {author} {\bibinfo {author} {\bibfnamefont {Paul~G}\ \bibnamefont
  {Nevai}},\ }\bibfield  {title} {\enquote {\bibinfo {title} {{G\'{e}za Freud,
  Orthogonal Polynomials. Christoffel Functions. A Case Study}},}\ }\href
  {\doibase 10.1016/0021-9045(86)90016-X} {\bibfield  {journal} {\bibinfo
  {journal} {Journal Of Approximation Theory}\ }\textbf {\bibinfo {volume}
  {48}},\ \bibinfo {pages} {3--167} (\bibinfo {year} {1986})}\BibitemShut
  {NoStop}%
\bibitem [{lap(2013)}]{lapack}%
  \BibitemOpen
  \href {http://www.netlib.org/lapack/} {\enquote {\bibinfo {title} {Lapack
  version 3.5.0},}\ } (\bibinfo {year} {2013})\BibitemShut {NoStop}%
\bibitem [{Note1()}]{Note1}%
  \BibitemOpen
  \bibinfo {note} {There are other ways to estimate thresholds. One can use
  e.g. Radau-like measures \hbox {$d\mathaccent "0365\relax {\mu }=(x-x_0)d\mu
  $} (or \hbox {$d\mathaccent "0365\relax {\mu }=(x_0-x)d\mu $} depending on
  $x_0$ position) or, another option, drop boundary condition (\protect \ref
  {psi0boundary}) altogether and estimate whether $I$ is ``low'' or ``high'' as
  closeness of $|\psi _{\protect \{IL,IH\protect \}}>$ to $|\psi _0>$,
  localized at $x_0$ (the one from (\ref {psix0norm})) as $<\psi _{\protect
  \{IL,IH\protect \}}|\psi _0>_{\mu }^2$. We will discuss this approach
  separately.}\BibitemShut {Stop}%
\bibitem [{Note2()}]{Note2}%
  \BibitemOpen
  \bibinfo {note} {The states corresponding to minimal $I$ poses similar
  properties, but as we noted in Section \protect \ref {EV} the $\psi _{IL}(x)$
  roots are simple real distinct (but not necessary on the support of the
  measure). Because $I$ is the lowest on this state, the high values of $I$
  should be localized near the $\psi _{IL}(x)$ roots). In most situations the
  results obtained near the $\psi _{IL}(x)$ roots are very similar to the
  calculations above on $|\psi _{IH}>$ state.}\BibitemShut {Stop}%
\bibitem [{\citenamefont {Taleb}(2010)}]{taleb2010black}%
  \BibitemOpen
  \bibfield  {author} {\bibinfo {author} {\bibfnamefont {Nassim~Nicholas}\
  \bibnamefont {Taleb}},\ }\href@noop {} {\emph {\bibinfo {title} {The black
  swan:: The impact of the highly improbable fragility}}},\ Vol.~\bibinfo
  {volume} {2}\ (\bibinfo  {publisher} {Random House},\ \bibinfo {year}
  {2010})\BibitemShut {NoStop}%
\bibitem [{\citenamefont {Laurie}\ and\ \citenamefont
  {Rolfes}(1979)}]{laurie1979computation}%
  \BibitemOpen
  \bibfield  {author} {\bibinfo {author} {\bibfnamefont {Dirk~P}\ \bibnamefont
  {Laurie}}\ and\ \bibinfo {author} {\bibfnamefont {Laurette}\ \bibnamefont
  {Rolfes}},\ }\bibfield  {title} {\enquote {\bibinfo {title} {{Computation of
  Gaussian quadrature rules from modified moments}},}\ }\href {\doibase
  10.1016/0377-0427(79)90008-6} {\bibfield  {journal} {\bibinfo  {journal}
  {Journal of Computational and Applied Mathematics}\ }\textbf {\bibinfo
  {volume} {5}},\ \bibinfo {pages} {235--243} (\bibinfo {year}
  {1979})}\BibitemShut {NoStop}%
\bibitem [{\citenamefont {Watson}(1938)}]{watson}%
  \BibitemOpen
  \bibfield  {author} {\bibinfo {author} {\bibfnamefont {G.N.}\ \bibnamefont
  {Watson}},\ }\bibfield  {title} {\enquote {\bibinfo {title} {{A note on the
  polynomials of Hermite and Laguerre}},}\ }\href {\doibase
  10.1112/jlms/s1-13.3.204} {\bibfield  {journal} {\bibinfo  {journal} {London
  Math. Soc. J.}\ }\textbf {\bibinfo {volume} {13}},\ \bibinfo {pages} {29}
  (\bibinfo {year} {1938})}\BibitemShut {NoStop}%
\bibitem [{\citenamefont {Gradshteyn}\ and\ \citenamefont
  {Ryzhik}(1963)}]{gradshtein}%
  \BibitemOpen
  \bibfield  {author} {\bibinfo {author} {\bibfnamefont {I.S.}\ \bibnamefont
  {Gradshteyn}}\ and\ \bibinfo {author} {\bibfnamefont {I.M.}\ \bibnamefont
  {Ryzhik}},\ }\href@noop {} {\emph {\bibinfo {title} {Table of Integrals,
  Series, and Products}}}\ (\bibinfo  {publisher} {Fizmatlit},\ \bibinfo {year}
  {1963})\BibitemShut {NoStop}%
\bibitem [{\citenamefont {Milne-Thomson}\ \emph {et~al.}(1972)\citenamefont
  {Milne-Thomson}, \citenamefont {Abramowitz},\ and\ \citenamefont
  {Stegun}}]{milne1972handbook}%
  \BibitemOpen
  \bibfield  {author} {\bibinfo {author} {\bibfnamefont {L~Melville}\
  \bibnamefont {Milne-Thomson}}, \bibinfo {author} {\bibfnamefont
  {M}~\bibnamefont {Abramowitz}}, \ and\ \bibinfo {author} {\bibfnamefont
  {IA}~\bibnamefont {Stegun}},\ }\href@noop {} {\emph {\bibinfo {title}
  {Handbook of mathematical functions}}}\ (\bibinfo  {publisher} {Dover
  Publications Nova Iorque},\ \bibinfo {year} {1972})\BibitemShut {NoStop}%
\bibitem [{\citenamefont {Carlitz}(1962)}]{Carlitz}%
  \BibitemOpen
  \bibfield  {author} {\bibinfo {author} {\bibfnamefont {L.}~\bibnamefont
  {Carlitz}},\ }\bibfield  {title} {\enquote {\bibinfo {title} {{The product of
  several Hermite or Laguerre polynomials}},}\ }\href {\doibase
  10.1007/BF01298234} {\bibfield  {journal} {\bibinfo  {journal} {Monatshefte
  fur Mathematik}\ }\textbf {\bibinfo {volume} {66(5)}},\ \bibinfo {pages}
  {393--396} (\bibinfo {year} {1962})}\BibitemShut {NoStop}%
\bibitem [{\citenamefont {Brychkov}\ \emph {et~al.}(2003)\citenamefont
  {Brychkov}, \citenamefont {Marichev},\ and\ \citenamefont
  {Prudnikov}}]{Prudnikov}%
  \BibitemOpen
  \bibfield  {author} {\bibinfo {author} {\bibfnamefont {Yu.~A.}\ \bibnamefont
  {Brychkov}}, \bibinfo {author} {\bibfnamefont {O.~I.}\ \bibnamefont
  {Marichev}}, \ and\ \bibinfo {author} {\bibfnamefont {A.~P.}\ \bibnamefont
  {Prudnikov}},\ }\href@noop {} {\emph {\bibinfo {title} {Integrals and Series.
  Volume 2}}}\ (\bibinfo {year} {2003})\BibitemShut {NoStop}%
\bibitem [{\citenamefont {Fox}\ and\ \citenamefont
  {Parker}(1968)}]{fox1968chebyshev}%
  \BibitemOpen
  \bibfield  {author} {\bibinfo {author} {\bibfnamefont {Leslie}\ \bibnamefont
  {Fox}}\ and\ \bibinfo {author} {\bibfnamefont {Ian~Bax}\ \bibnamefont
  {Parker}},\ }\href@noop {} {\emph {\bibinfo {title} {Chebyshev polynomials in
  numerical analysis}}},\ Vol.~\bibinfo {volume} {29}\ (\bibinfo  {publisher}
  {Oxford university press London},\ \bibinfo {year} {1968})\BibitemShut
  {NoStop}%
\bibitem [{\citenamefont {Maroulas}\ and\ \citenamefont
  {Barnett}(1979)}]{Barnett}%
  \BibitemOpen
  \bibfield  {author} {\bibinfo {author} {\bibfnamefont {John}\ \bibnamefont
  {Maroulas}}\ and\ \bibinfo {author} {\bibfnamefont {Stephen}\ \bibnamefont
  {Barnett}},\ }\bibfield  {title} {\enquote {\bibinfo {title} {{Polynomials
  With Respect to a General Basis. II. Applications}},}\ }\href {\doibase
  10.1016/0022-247X(79)90251-8} {\bibfield  {journal} {\bibinfo  {journal}
  {Journal of Matematical Alalysis Applications}\ }\textbf {\bibinfo {volume}
  {72}},\ \bibinfo {pages} {599--614} (\bibinfo {year} {1979})}\BibitemShut
  {NoStop}%
\bibitem [{\citenamefont {T.Bella}\ \emph {et~al.}(2006)\citenamefont
  {T.Bella}, \citenamefont {Y.Eidelman}, \citenamefont {I.Gohberg},
  \citenamefont {V.Olshevsky},\ and\ \citenamefont {E.Tyrtyshnikov}}]{Bella}%
  \BibitemOpen
  \bibfield  {author} {\bibinfo {author} {\bibnamefont {T.Bella}}, \bibinfo
  {author} {\bibnamefont {Y.Eidelman}}, \bibinfo {author} {\bibnamefont
  {I.Gohberg}}, \bibinfo {author} {\bibnamefont {V.Olshevsky}}, \ and\ \bibinfo
  {author} {\bibnamefont {E.Tyrtyshnikov}},\ }\bibfield  {title} {\enquote
  {\bibinfo {title} {Fast inversion of hessenberg-quasiseparable-vandermonde
  matrices and resulting recurrence relations and characterizations},}\ }\href
  {http://www.math.uconn.edu/~olshevsky/papers/traubqs2_tb36.pdf} {\bibfield
  {journal} {\bibinfo  {journal} {preprint}\ } (\bibinfo {year}
  {2006})}\BibitemShut {NoStop}%
\bibitem [{\citenamefont {Malyshkin}(2014)}]{polynomialcode}%
  \BibitemOpen
  \bibfield  {author} {\bibinfo {author} {\bibfnamefont
  {Vladislav~Gennadievich}\ \bibnamefont {Malyshkin}},\ }\href
  {http://www.ioffe.ru/LNEPS/malyshkin/code.html} {} (\bibinfo {year} {2014}),\
  \bibinfo {note} {the code for polynomials calculation,
  \url{http://www.ioffe.ru/LNEPS/malyshkin/code.html}}\BibitemShut {NoStop}%
\bibitem [{\citenamefont {Kronrod}(1964)}]{kronrod1964}%
  \BibitemOpen
  \bibfield  {author} {\bibinfo {author} {\bibfnamefont {Aleksandr~Semenovich}\
  \bibnamefont {Kronrod}},\ }\href@noop {} {\emph {\bibinfo {title} {Nodes and
  Weights of Quadratures formulas: 16-digits tables (in Russian)}}}\ (\bibinfo
  {publisher} {Nauka},\ \bibinfo {year} {1964})\BibitemShut {NoStop}%
\bibitem [{\citenamefont {Laurie}(1997)}]{laurie1997calculation}%
  \BibitemOpen
  \bibfield  {author} {\bibinfo {author} {\bibfnamefont {Dirk}\ \bibnamefont
  {Laurie}},\ }\bibfield  {title} {\enquote {\bibinfo {title} {{Calculation of
  Gauss-Kronrod quadrature rules}},}\ }\href {\doibase
  10.1090/S0025-5718-97-00861-2} {\bibfield  {journal} {\bibinfo  {journal}
  {Mathematics of Computation of the American Mathematical Society}\ }\textbf
  {\bibinfo {volume} {66}},\ \bibinfo {pages} {1133--1145} (\bibinfo {year}
  {1997})}\BibitemShut {NoStop}%
\bibitem [{\citenamefont {Borges}(1994)}]{borges1994class}%
  \BibitemOpen
  \bibfield  {author} {\bibinfo {author} {\bibfnamefont {Carlos~F}\
  \bibnamefont {Borges}},\ }\bibfield  {title} {\enquote {\bibinfo {title} {{On
  a class of Gauss-like quadrature rules}},}\ }\href {\doibase
  10.1007/s002110050028} {\bibfield  {journal} {\bibinfo  {journal} {Numerische
  Mathematik}\ }\textbf {\bibinfo {volume} {67}},\ \bibinfo {pages} {271--288}
  (\bibinfo {year} {1994})}\BibitemShut {NoStop}%
\bibitem [{\citenamefont {Szeg\H{o}}(1939)}]{szego1974orthogonal}%
  \BibitemOpen
  \bibfield  {author} {\bibinfo {author} {\bibfnamefont {Gabor}\ \bibnamefont
  {Szeg\H{o}}},\ }\href@noop {} {\emph {\bibinfo {title} {Orthogonal
  Polynomials}}},\ \bibinfo {edition} {fourth edition, 1975}\ ed.,\ \bibinfo
  {series} {American Mathematical Society colloquium publications},
  Vol.~\bibinfo {volume} {23}\ (\bibinfo  {publisher} {American Mathematical
  Society},\ \bibinfo {year} {1939})\BibitemShut {NoStop}%
\bibitem [{\citenamefont {von Hippel}(2005)}]{von2005mean}%
  \BibitemOpen
  \bibfield  {author} {\bibinfo {author} {\bibfnamefont {Paul~T}\ \bibnamefont
  {von Hippel}},\ }\bibfield  {title} {\enquote {\bibinfo {title} {Mean,
  median, and skew: Correcting a textbook rule},}\ }\href
  {http://www.amstat.org/publications/JSE/v13n2/vonhippel.html} {\bibfield
  {journal} {\bibinfo  {journal} {Journal of Statistics Education}\ }\textbf
  {\bibinfo {volume} {13}} (\bibinfo {year} {2005})}\BibitemShut {NoStop}%
\bibitem [{\citenamefont {Malyshkin}(2015)}]{malyshkin2015norm}%
  \BibitemOpen
  \bibfield  {author} {\bibinfo {author} {\bibfnamefont
  {Vladislav~Gennadievich}\ \bibnamefont {Malyshkin}},\ }\bibfield  {title}
  {\enquote {\bibinfo {title} {{Norm-Free Radon-Nikodym Approach to Machine
  Learning}},}\ }\href {http://arxiv.org/abs/1512.03219} {\bibfield  {journal}
  {\bibinfo  {journal} {ArXiv e-prints}\ } (\bibinfo {year} {2015})},\ \bibinfo
  {note} {\url{http://arxiv.org/abs/1512.03219}},\ \Eprint
  {http://arxiv.org/abs/1512.03219} {arXiv:1512.03219 [cs.LG]} \BibitemShut
  {NoStop}%
\end{thebibliography}%

\end{document}